\documentclass[amsmath,amssymb,twocolumn]{revtex4}
\usepackage{dcolumn}
\usepackage{bm}
\usepackage{amsmath,mathrsfs}
\usepackage{graphicx,epsfig}
\usepackage{hyperref}
\usepackage{epsf}
\usepackage{color}

\begin{document}


\title{{\bf Traversable wormholes satisfying energy conditions in $f(Q)$ gravity }}

\author{S. Rastgoo}\email{rastgoo@sirjantech.ac.ir}
\author{ F. Parsaei }\email{fparsaei@gmail.com}

\affiliation{ Physics Department , Sirjan University of Technology, Sirjan 78137, Iran.}

\date{\today}


\begin{abstract}
\par   In this article, a new family of asymptotically flat wormhole solutions in the context of symmetric teleparallel gravity, i.e., $f(Q)$ theory of gravity, are presented. Considering a power-law shape function and some different forms for $f(Q)$ function, we show that a wide variety of wormhole solutions for which the matter fields satisfy some  energy conditions, are accessible. We explore that the presence of $f(Q)$ gravity will be enough to sustain a traversable wormhole without exotic matter. The influence of free parameters in shape function and $f(Q)$ models on the energy conditions is investigated. The equation of state and boundary conditions are analyzed.  \\
\end{abstract}

\maketitle
\section{Introduction}
Traversable wormholes are solutions of the Einstein’s field equations with exotic sources. A traversable wormhole solution should not have any horizon or singularity \cite{Visser}. Theoretically, a wormhole provides a short way between the points in the different universes or two points in the same universe. Time machine may be a consequence of the existence of wormholes. The first proposal of a wormhole was presented by flamm \cite{flamm}  then in 1935 Einstein and Rosen constructed the Einstein-Rosen bridge \cite{Rosen} which is basically the maximally extended Schwarzschild solution. It should be noted that the Einstein-Rosen bridge is not a traversable wormhole. Historically, Misner and Wheeler have introduced the term “wormhole” in 1957  \cite{wheeler}  but  the general basic structure of the traversable  wormhole is presented   by Morris and Thorne \cite{WH}.
 It was shown that wormholes need exotic matter to be constructed \cite{Visser}. So in the context of general relativity (GR), violation of null energy condition (NEC) i.e.  $T_{\mu\nu}k^{\mu}k^{\nu}\geq0$, in which $k^{\mu}$ is any null vector and $T_{\mu\nu}$ stress-energy tensor, is the main ingredients of the wormhole theory.

 Distance measurements of type Ia supernovae demonstrated accelerated expansion of the Universe which is investigated extensively in the literature. In this view, two main approaches are used to explain the  accelerated expansion of the Universe. The first one explains this phenomena, by introducing a new field in the context of GR, which is famous as dark energy (DE). Fluid with an equation of state (EoS), $p=\omega \rho $, and positive energy density  is a good candidate to explain the evolution of the cosmos. The DE admits $-1<\omega\leq 0$ while  $\omega\leq -1$ is defined as the phantom regime. Fluid  with $\omega\leq -\frac{1}{3}$, may be considered as a source to describe  accelerated expansion of the Universe.  It should be noted that phantom fluid violates NEC so the wormholes with phantom EoS  are studied extensively in the literature \cite{phantom, phantom1}.

 One can minimize the amount of exotic material, by constructing a thin-shell wormhole \cite{cut} or finding the wormhole with variable EoS \cite{Remo, variable}. In thin-shell wormhole, the exotic matter is confined near the throat of wormhole. On the other side, in \cite{foad}, it was shown that wormholes with polynomial EoS in contrast to linear EoS  violate the energy conditions (ECs) only in a small region of the spacetime.

 In the second approach, alternative theories of gravity have been used to study the  accelerated expansion of the Universe. Usually, the generalization of the wormhole by alternative theory is another way to solve the problem of exotic matter. Modeling traversable wormholes in the framework of modified gravity is an interesting task in the wormhole theory.  Several proposals have been presented to explain the exotic matter in the wormhole theory, by using modified theories of gravity.  Braneworld \cite{brane}, Born-Infeld theory \cite{Born}, quadratic gravity \cite{quad}, Einstein-Cartan gravity \cite{Cartan},  hybrid metric-Palatini gravity,  $f(R)$ gravity\cite{f(R)} and $f(R,T)$ gravity \cite{f(RT), f(T)}  are some examples. In most of these theories, the right-hand side of Einstein's field equations is modified. This modification provides the conditions to construct wormhole solutions without exotic matter or at least minimizing  the usage of exotic matter.

The $f(Q)$ gravity is a consequence of  constructing new classes of modified gravity, starting from the symmetric teleparallel gravity, which is based on the non-metricity scalar $Q$. In fact, the $f(Q)$ gravity is a generalization of the symmetric teleparallel gravity which is organized on a flat and vanishing
torsion connection. Jimenez et al. \cite{Jim} have proposed a class of theory in which  both curvature and torsion vanish, and gravity is attributed to non-metricity ($Q$).
The $f(Q)$ theory can  describe the  accelerated expansion of the Universe at least to the same level of statistic precision of most renowned modified gravities \cite{Zha}.
The cosmology in $f(Q)$ gravity has been studied in \cite{Jim2, Tel, Tel2}.
Harko et al. have used $f(Q)$ gravity to describe cosmological evolutions and other aspects \cite{Tel}. They have shown that  an extension of the
symmetric teleparallel gravity, by considering a new class of theories where the nonmetricity  is coupled nonminimally to the matter Lagrangian, in the framework of
the metric-affine formalism, can provide gravitational alternatives to DE. Some researchers have studied the ECs \cite{Ec}  and Newtonian limit \cite{Newt} in the background  of $f(Q)$ gravity.  Anagnostopoulo et al. have used  the Big Bang Nucleosynthesis  formalism and observations  to extract constraints on various classes of $f(Q)$ models \cite{Big}. Many other researchers have investigated the cosmology in $f(Q)$ gravity \cite{cosm}.

Besides, black hole solutions have been investigated  in the context of $f(Q)$ gravity \cite{Black, Black2}.
Also, the compact star generated by gravitational decoupling in $f(Q)$ gravity theory is studied in \cite{Star}. Sokoliuk et al. \cite{Star2} have explored the Buchdahl quark stars in the background of $f(Q)$ theory. Recently, wormhole and spherically symmetric configurations are studied in $f(Q)$ gravity \cite{Zha} and \cite{ Wang, Sha, Mus, Hassan, sym, Ban, Calz, f(Q), Kir}. In  \cite{Zha} a simple model, $f(Q)=Q+\alpha Q^{2}$, with a polytropic EoS is considered to find the internal spherically symmetric configuration.
 The static and spherically symmetric solutions with an anisotropic fluid for general $f(Q)$ gravity are presented by Wang et al. \cite{Wang}. Sharma et al. have studied the wormhole solutions in the context of symmetric teleparallel gravity \cite{Sha}. They have shown solutions with special shape and redshift functions for some of the $f(Q)$ models can provide solutions that satisfy ECs in some regions of the spacetime.
Mustafa et al. have used the Karmarkar condition in $f(Q)$ gravity formalism to find  wormhole solutions which satisfy the ECs \cite{Mus}. In \cite{Hassan}, the traversable wormhole geometries in $f(Q)$ by considering two specific EoS are investigated. The solutions in \cite{Hassan} have been obtained for a specific
shape function in the fundamental interaction of gravity (i.e., for a linear form of $f(Q)$). This class of solutions do not respect the ECs.
In \cite{sym}, the authors discussed the existence of wormhole solutions with the help of the Gaussian and Lorentzian distributions of linear and exponential models. They have shown that the wormhole solutions obtained with these models are physically capable and stable but do not respect the ECs.
Solutions with constant  redshift function and different shape functions are presented in \cite{Ban} for some known $f(Q)$ functions. This class of solutions  violate the ECs. Along this way, Calza and  Sebastiani have analyzed a class of topological static spherically symmetric vacuum solutions in $f(Q)$ gravity with constant non-metricity \cite{Calz}. In\cite{f(Q)}, we have studied the wormhole in the background of $f(Q)$ gravity. The solutions are presented by  focusing on $f(Q)$ gravity, introduced by Jimenez et al.,\cite{Jim}. We have found solutions which violate the ECs only in some small regions of the spacetime. In this work, we are particularly interested in
finding some specific traversable wormhole solutions without requiring any exotic matter.

 The organization of the paper is as follows: According to \cite{f(Q)}, first, we discussed conditions and equations governing wormhole then a brief review of $f(Q)$ theory and the classical ECs is presented. In Sec. \ref{sec3} by defining a known shape function, we have analyzed  some basic $f(Q)$ functions to find  solutions which can satisfy the ECs. The physical properties of the solutions that satisfy the ECs are presented in this section. Finally, we have presented our concluding remarks in the last section. In this paper, We have assumed gravitational units, i.e., $c = 8 \pi G = 1$.

\section{Basic formulation of wormhole }
In this section, we will try to introduce the basic structure of the wormhole theory and a brief review of $f(Q)$ gravity formalism. We use the prescription introduced in
Ref.  \cite{f(Q)} where a detailed discussion about the formulation of the theory can be found.
 We use the line element of the general spherically symmetric wormhole as:
\begin{equation}\label{1}
ds^2=-U(r)dt^2+\frac{dr^2}{1-\frac{b(r)}{r}}+r^2(d\theta^2+\sin^2\theta,
d\phi^2)
\end{equation}
where $U(r)=\exp (2\phi(r))$. The function $b(r)$ and $ \phi(r)$ are functions of the radial coordinate. The former is called the redshift function which can be used to detect the redshift of the signal  by a distance observer. Also $b(r)$  is called the shape or form function. The shape function should obey the condition

\begin{equation}\label{2}
b(r_0)=r_0.
\end{equation}
where  $r_0$ is the wormhole throat.  Two other conditions are imposed as follows to have a traversable wormhole,

\begin{equation}\label{3}
b'(r_0)<1
\end{equation}
and
\begin{equation}\label{4}
b(r)<r,\ \ {\rm for} \ \ r>r_0.
\end{equation}
The former is famous as flaring out condition. In the classical GR, the flaring-out condition  and the NEC  are incompatible.
In this paper, we will  consider asymptotically flat condition as follows
\begin{equation}\label{5}
\lim_{r\rightarrow \infty}\frac{b(r)}{r}=0,\qquad   \lim_{r\rightarrow \infty}U(r)=1
\end{equation}

It is worth mentioning that constant redshift function guarantees the absence of horizon around the throat and presents zero tidal force. Physically, wormhole solutions with
constant or non-constant redshift function do not have much difference, so for the sake of simplicity, we have considered solutions with constant redshift function.

Let us briefly review the $f(Q)$ formalism. The action for  symmetric teleparallel gravity is given by
\begin{equation}\label{6}
S=\int\frac{1}{2}f(Q)\sqrt{-g}\;d^{4}x+\int L_{m}\sqrt{-g}\;d^{4}x.
\end{equation}
where $f(Q)$ is a general function  of $Q$, $g$ is the determinant
of the metric, and $L_m$ is the matter Lagrangian density. Now, one can define the non-metricity
tensor and its trace  by
\begin{equation}\label{7}
Q_{\lambda\mu\nu}=\nabla_{\lambda}g_{\mu\nu},
\end{equation}
\begin{equation}\label{8}
Q_{\alpha}=Q_{\alpha\mu}^{\mu}\;\;,\;\;\widetilde{Q}_{\alpha}=Q^{\mu}_{\alpha\mu}.
\end{equation}
Further, the non-metricity conjugate is presented by
\begin{equation}\label{9}
P^{\alpha}_{\mu\nu}=\frac{1}{4}[-Q^{\alpha}_{\mu\nu}+2Q_{\mu}^{\alpha}+Q^{\alpha}g_{\mu\nu}-\widetilde{Q}^{\alpha}g_{\mu\nu}-\delta^{\alpha}_{\mu}Q_{\nu}]
\end{equation}
so
\begin{equation}\label{10}
Q=-Q_{\alpha\mu\nu}P^{\alpha\mu\nu}.
\end{equation}
On the other hand,  the energy-momentum tensor is shown by
\begin{equation}\label{11}
T_{\mu\nu}=-\frac{2}{\sqrt{-g}}\frac{\delta(\sqrt{-g}L_{m})}{\delta\;g^{\mu\nu}}.
\end{equation}
In this realm, the field equations  are obtained by varying the action (\ref{6}) with respect to the metric

\begin{eqnarray}\label{12}
T_{\mu\nu}=-\frac{2}{\sqrt{-g}}\nabla_{\gamma}(\sqrt{-g}f_{Q}P^{\gamma}_{\mu\nu})-\frac{1}{2}g_{\mu\nu}f \nonumber \\
-f_{Q}(P_{\mu\gamma i}Q_{\nu}^{\gamma i}-2Q_{\gamma i\mu}P^{\gamma i}_{\nu}).
\end{eqnarray}
and
\begin{equation}\label{13}
\nabla_{\mu}\nabla_{\nu}(\sqrt{-g}f_{Q}P^{\gamma}_{\mu\nu})=0.
\end{equation}
 where $f_Q\equiv \frac{df}{dQ}$. The metric (\ref{1}) leads to the non-metricity tensor
\begin{equation}\label{14}
Q=-\frac{2}{r}(1-\frac{b(r)}{r})(2\phi^{\prime}+\frac{1}{r}).
\end{equation}
 We shall assume that matter is well described by an anisotropic perfect fluid, i.e., the stress-energy tensor
 can be written in the form  $T^{\mu}_{\nu}=diag[-\rho, p,p_t,p_t]$, where $\rho$ is the energy density, $p$ the
 radial pressure and $p_t$ the tangential pressure, respectively. Using (\ref{1}), (\ref{14}) and (\ref{12}), one can find the following field equations
\begin{eqnarray}\label{15}
\rho=&[&\frac{1}{r}(\frac{1}{r}-\frac{rb^{\prime}(r)+b(r)}{r^{2}}+2\phi^{\prime}(r)(1-\frac{b(r)}{r})]f_{Q} \nonumber \\
&+&\frac{2}{r}(1-\frac{b(r)}{r})f'_{Q}+\frac{f}{2}.
\end{eqnarray}

\begin{equation}\label{16}
p_{r}=-[\frac{2}{r}(1-\frac{b(r)}{r})(2\phi^{\prime}(r)+\frac{1}{r})-\frac{1}{r^{2}}]f_{Q}-\frac{f}{2}.
\end{equation}

\begin{eqnarray}\label{17}
p_t(r)=[\frac{1}{r}((1-\frac{b(r)}{r})(\frac{1}{r}+\phi^{\prime}(r)(3+r\phi^{\prime}(r))\nonumber \\
+r\phi^{\prime\prime})-\frac{rb^{\prime}(r)
 -b(r)}{2r^{2}}(1+r\phi^{\prime}(r)))]f_{Q}\nonumber \\
+\frac{1}{r}((1-\frac{b(r)}{r}))(1+r\phi^{\prime}(r))f_{QQ}+\frac{f}{2}.
\end{eqnarray}
which the prime denotes the derivative $\frac{d}{dr}$. Now, we have the essential mathematical tools to study the wormhole solutions in the background of $f(Q)$. As  was mentioned in the introduction, many algorithms have been used to find wormhole solutions in the $f(Q)$ but we will use a known shape function and some $f(Q)$ models with free parameters to explore new wormhole solutions.

One of the main ingredients of wormhole solutions in the ordinary GR is the violation of ECs.  The ECs represent paths to accomplish the positiveness of the stress-energy tensor in the presence of matter. The four known ECs which are famous as the  null energy condition (NEC), dominant energy condition (DEC), weak energy condition (WEC), and strong energy condition (SEC), are defined as:
\begin{eqnarray}\label{18}
\textbf{NEC}&:& \rho+p\geq 0,\quad \rho+p_t\geq 0 \\
\label{19}
\textbf{WEC}&:& \rho\geq 0, \rho+p\geq 0,\quad \rho+p_t\geq 0, \\
\textbf{DEC}&:& \rho\geq 0, \rho-|p|\geq 0,\quad \rho-|p_t|\geq 0, \\
\textbf{SEC}&:& \rho+p\geq 0,\, \rho+p_t\geq 0,\rho+p+2p_t \geq 0. \label{20}
\end{eqnarray}

These conditions are the essential tools to understand the geodesics of the Universe which  can be derived from the well-known Raychaudhury equations. The ECs can be used to explain the attractive nature of gravity, besides assigning the fundamental causal and the geodesic structure of spacetime \cite{Capo}. According to \cite{f(Q)}, by defining the functions,
\begin{eqnarray}\label{21}
 H(r)&=& \rho+p ,\, H_1(r)= \rho+p_t,\, H_2(r)= \rho-|p|, \nonumber \\
 H_3(r)&=&\rho-|p_t|,\, H_4(r)= \rho+p+2p_t ,
\end{eqnarray}
we can investigate the ECs in the recent part of this paper.

In this article, we will focus on finding solutions in the background of $f(Q)$ gravity that satisfy ECs. To have some physically viable and reasonable models of traversable wormholes and to discover  the possibility of non-exotic wormholes within the framework of $f(Q)$ gravity, in the next parts of the paper, we will try to consider a known shape function with some $f(Q)$ models. We will see that the existence of free parameters in the shape function and $f(Q)$ functions provides solutions which respect the ECs.

\section{Wormhole solutions }\label{sec3}

 There are several motivations to explore wormhole in theories beyond the standard formulation of gravity. The most important one is solving the problem of exotic matter. As we know, solutions that respect ECs, without any extra condition, in the background of $f(Q)$ gravity introduced by Jimenez et al.  \cite{Jim} are not presented yet. In this section, we will try to find solutions which respect the ECs. For the sake of simplicity, we set  $r_0 =1$ and $\phi(r)=0$. By using equations (\ref{15})-(\ref{17}), and functions, namely, $f(Q)$, $b(r)$, $\phi(r)$, the energy momentum tensor will be explored. Usually, an EoS is added to the field equations as the extra condition and one of the functions $f(Q)$, $b(r)$, and $\phi(r)$ is considered unknown then one can find the unknown function through the field equations. The complexity of equations in the context of $f(Q)$ gravity does not allow us to use this algorithm for finding wormhole solutions that satisfy ECs. We will use known functions with free parameters then by fine-tuning the free parameters, we will  construct a wormhole with the most consistency. Along this way, we  consider a known shape function
  \begin{equation}\label{24}
b(r)=r^{m}.
\end{equation}
This shape function is the most famous one in the wormhole theory. It satisfies all of the necessary conditions  to construct a traversable wormhole \cite{phantom1}. It is easy to show that asymptotically flat condition (\ref{4}) implies that $m<1$ is acceptable. We will use some different forms of the $f(Q)$ function with this shape function to achieve wormhole solutions which respect the ECs. Equation(\ref{14}) along with shape function (\ref{24}) leads to
 \begin{equation}\label{24a1}
Q(r)= -2(r^{-2}-r^{m-3})
\end{equation}
In the next sections, by using some different models of $f(Q)$, we will seek solutions that satisfy the ECs. We start with some general forms of the $f(Q)$ function to construct the desired wormhole solutions then we will study the physical properties of the solutions.

\begin{figure}
\centering
  \includegraphics[width=3 in]{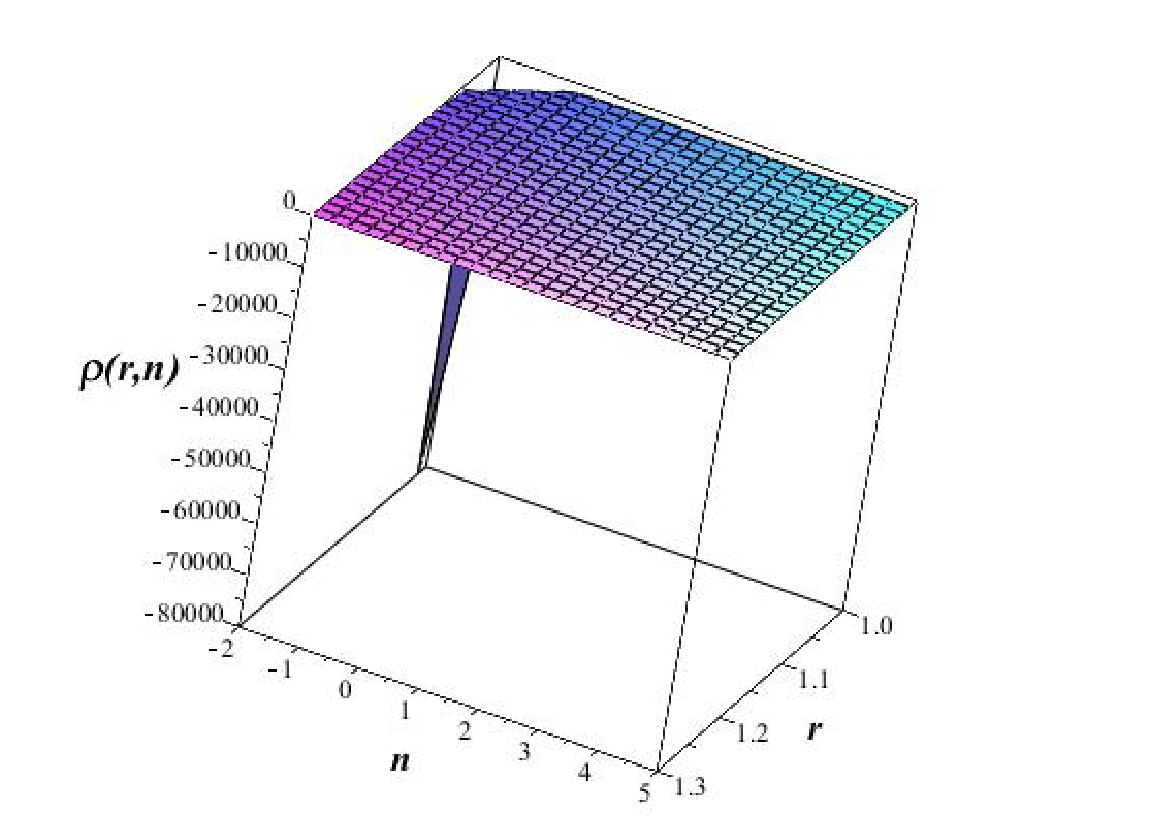}
\caption{The figure represents the $\rho(r,n)$ against radial coordinate and $n$ for $m=-2$,  which is  negative in the whole range. See the text for details.}
 \label{fig1}
\end{figure}

\begin{figure}
\centering
  \includegraphics[width=3 in]{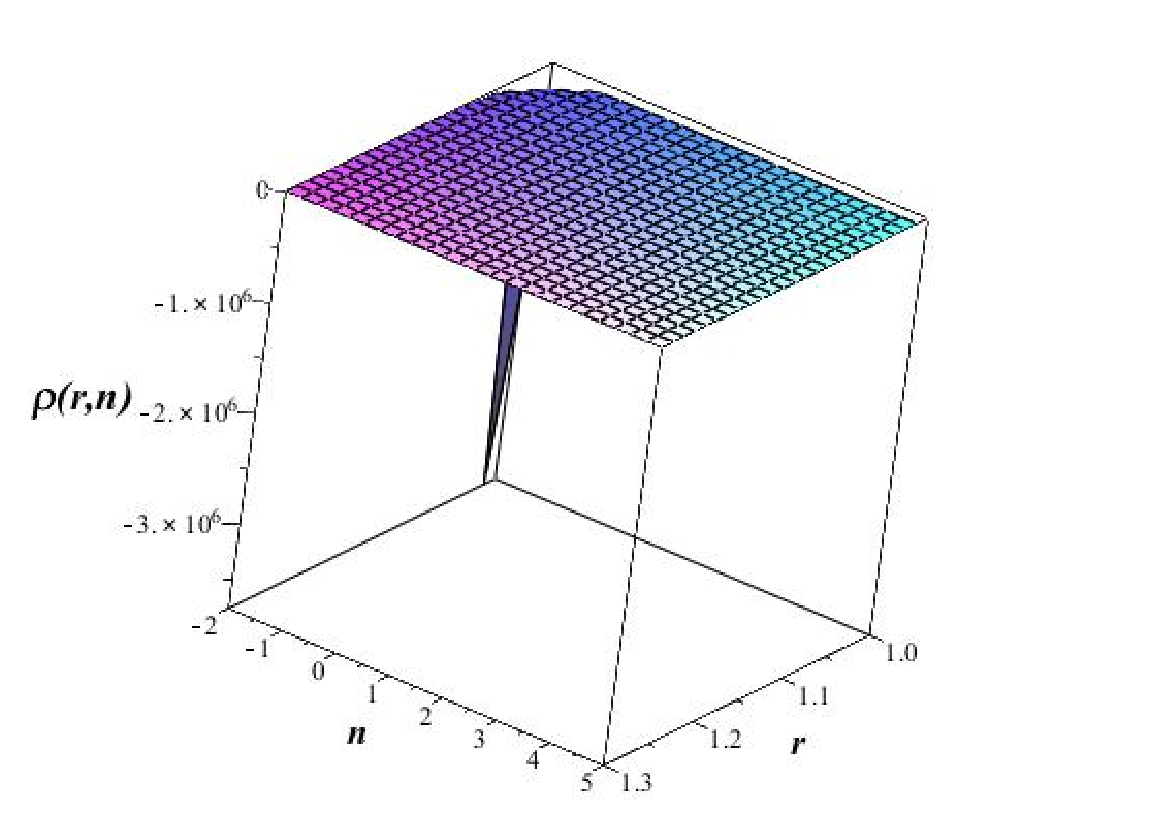}
\caption{The figure represents the $\rho(r,n)$ against radial coordinate and $n$ for $m=1/2$,  which is  negative in the whole range. See the text for details.}
 \label{fig2}
\end{figure}
\begin{figure}
\centering
  \includegraphics[width= 3 in]{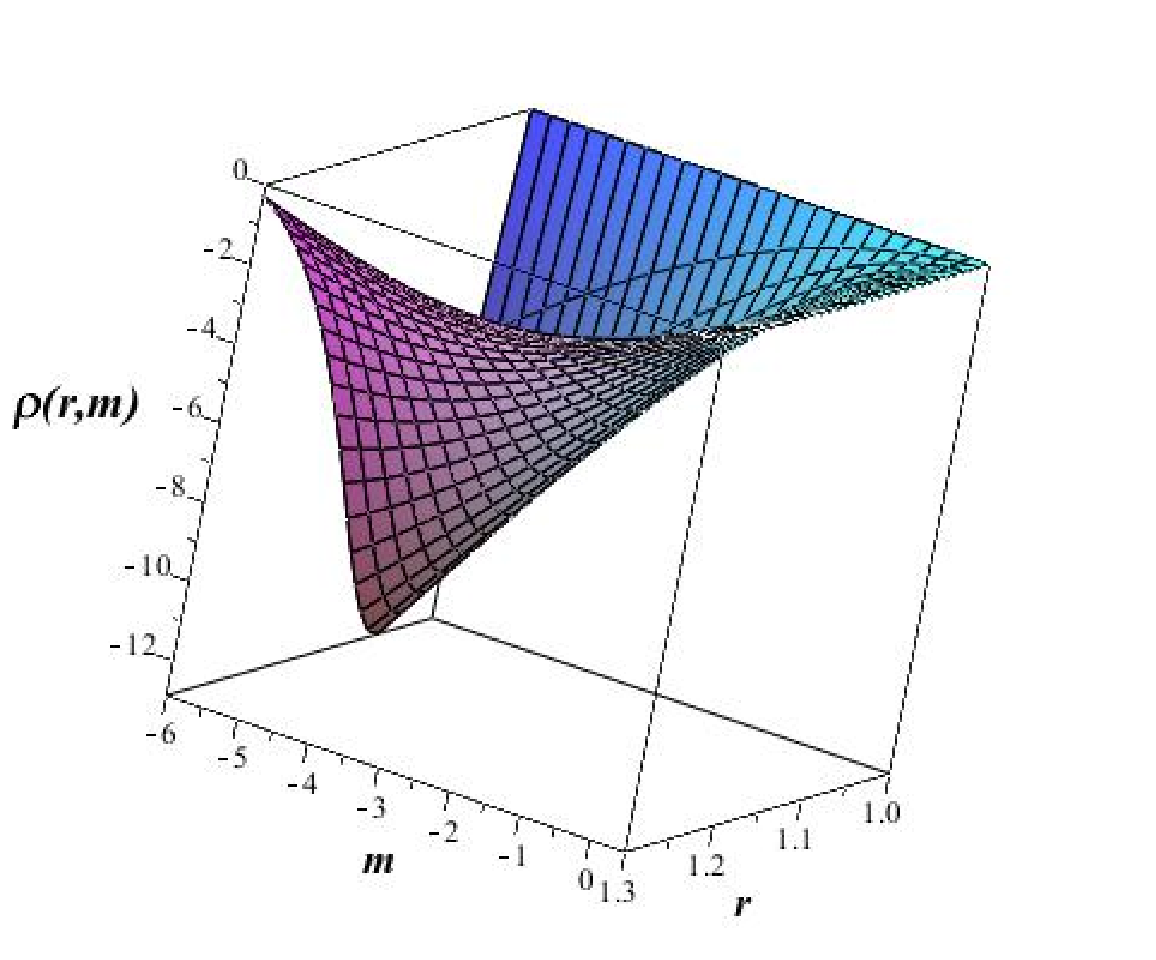}
\caption{The plot depicts the general behavior of $\rho(r,m)$ against $r$  and $m$ for the case  $n=2$. It is clear that $\rho(r,m)$ is  negative in the entire range of $r$ and $m$. It shows that the ECs are violated. See the text for details.}
 \label{fig3}
\end{figure}

\subsection{$f(Q)=(-Q)^{n}$}\label{subsec1}
The function
\begin{equation}\label{24a}
f(Q)=(-Q)^{n}
\end{equation}
is the first candidate to investigate ECs in the background of $f(Q)$ scenario. The general form of this  function has been studied in the context of $f(Q)$ in \cite{Tel}. It is clear that studying solutions with form function (\ref{24}) and a general form of (\ref{24a}) is very complicated. For the sake of simplicity, we use the shape function (\ref{24a}) with some special value for $m$ i.e., $m=1/2,0, -1/2, -1, -2, -4$. We have plotted the energy density function for some of these choices in Figs. (\ref{fig1}) and (\ref{fig2}). These figures imply that the energy density for this kind of solutions is negative so this class of solutions is not considerable. Although the general form function $b(r)=r^{m}$ with arbitrary $m$ is not examined but the general behavior of the energy density diagram for some different value of  $m$ corroborate the phrase that this class of solutions can not satisfy the ECs.

 In the second case, we consider a $f(Q)$ function with constant $n$ and variable $m$ for the shape function. The energy density is plotted for $n=2$ in the Fig.(\ref{fig3}) as a function of $m$ and radial coordinate. This figure demonstrates that $\rho$ is negative so this class of solutions  is not significant. One can use this algorithm for different values of $n$ but it seems that the result will be the same. As it was seen, the positive energy density is not accessible with this class of shape and $f(Q)$ functions so in the next subsection another $f(Q)$ model is examined.

\subsection{$f(Q)=Q^{2}+c$ }\label{subsec2}
We will continue our study with a $f(Q)$ function in the form
 \begin{eqnarray}\label{25}
f(Q)=Q^{2}+c .
\end{eqnarray}
It is a second-order function of $Q$ which a constant parameter is added. One can show that
\begin{figure}
\centering
  \includegraphics[width=3 in]{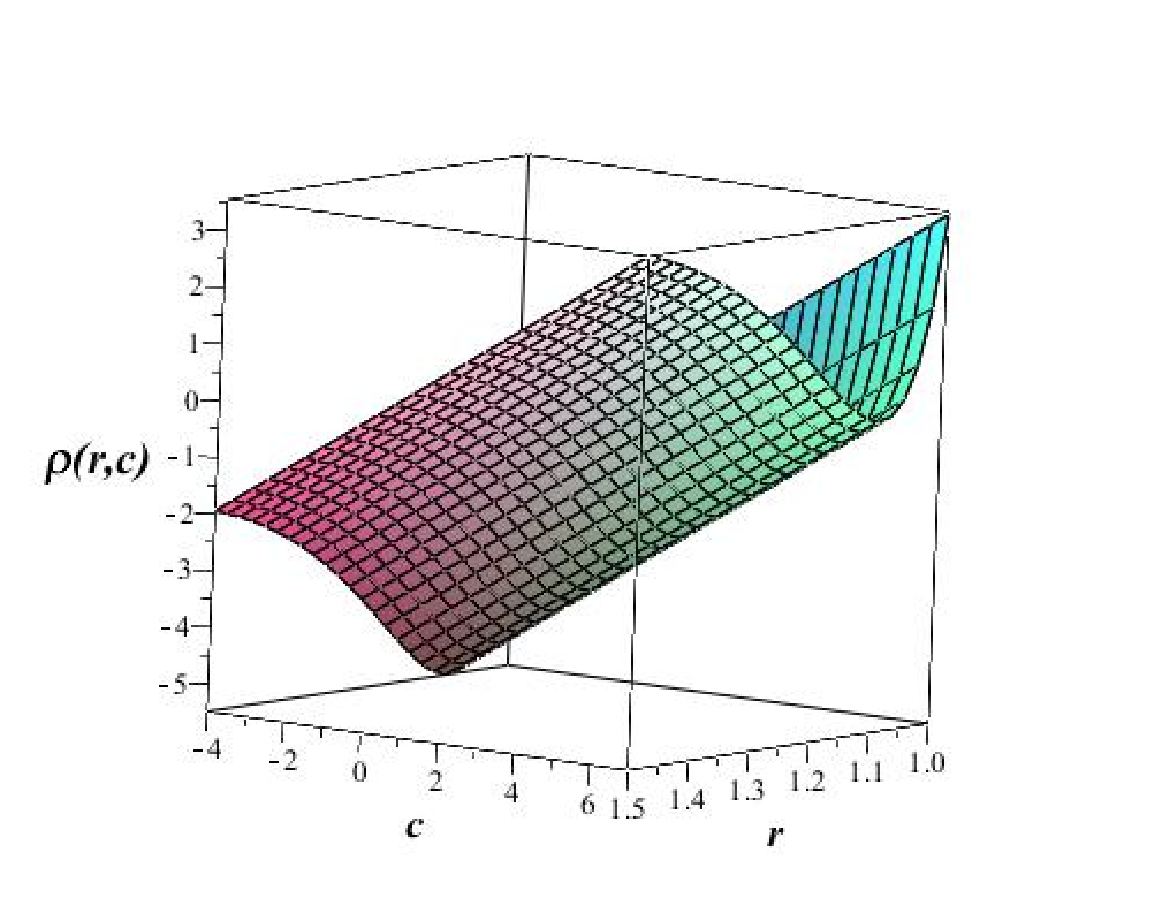}
\caption{The figure represents the $\rho(r,c)$ against radial coordinate and $c$ for $m=-2$, which is positive for some range. See the text for details.}
 \label{fig4}
\end{figure}

 \begin{equation}\label{26}
\lim_{r\rightarrow \infty}\rho = \frac{c}{2}
\end{equation}
This implies that the constant parameter $c$ can be set as $c=2 \rho_{\infty}$ which $\rho_{\infty}$ is the energy density at large distance. So the positive value of $c$ is considerable. We have plotted $\rho(r,c)$ as a function of $r$ and $c$ for $m=-2$  in Fig.(\ref{fig4}) which shows that the energy density is positive for some range of $c$. For the next step, we have plotted the $H(r,c)=\rho+p$ in the Fig.(\ref{fig5}). It is clear that  the ECs are not satisfied in this class of solutions. The same results can be concluded for the shape function with values $m=1/2, -1/2, -1, -2, -3, -4$.

\begin{figure}
\centering
    \includegraphics[width=3 in]{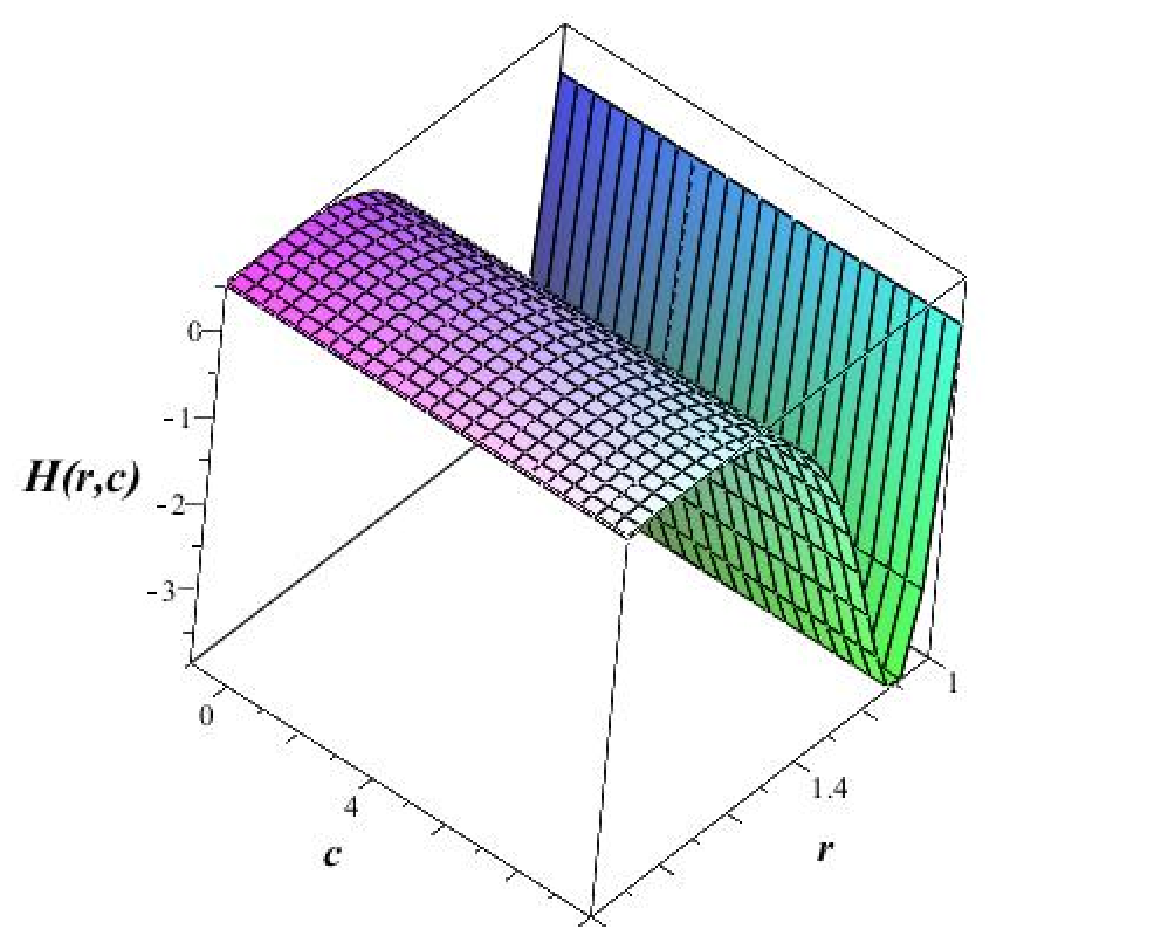}
\caption{The graphical behavior of $H(r,c)$,  against $r$ and $c$ for the case $m=-2$. It is clear that this function is negative in the entire range of $r$ and $c$ so the ECs are violated. See the text for details.}
 \label{fig5}
\end{figure}

\begin{figure}
\centering
  \includegraphics[width=3 in]{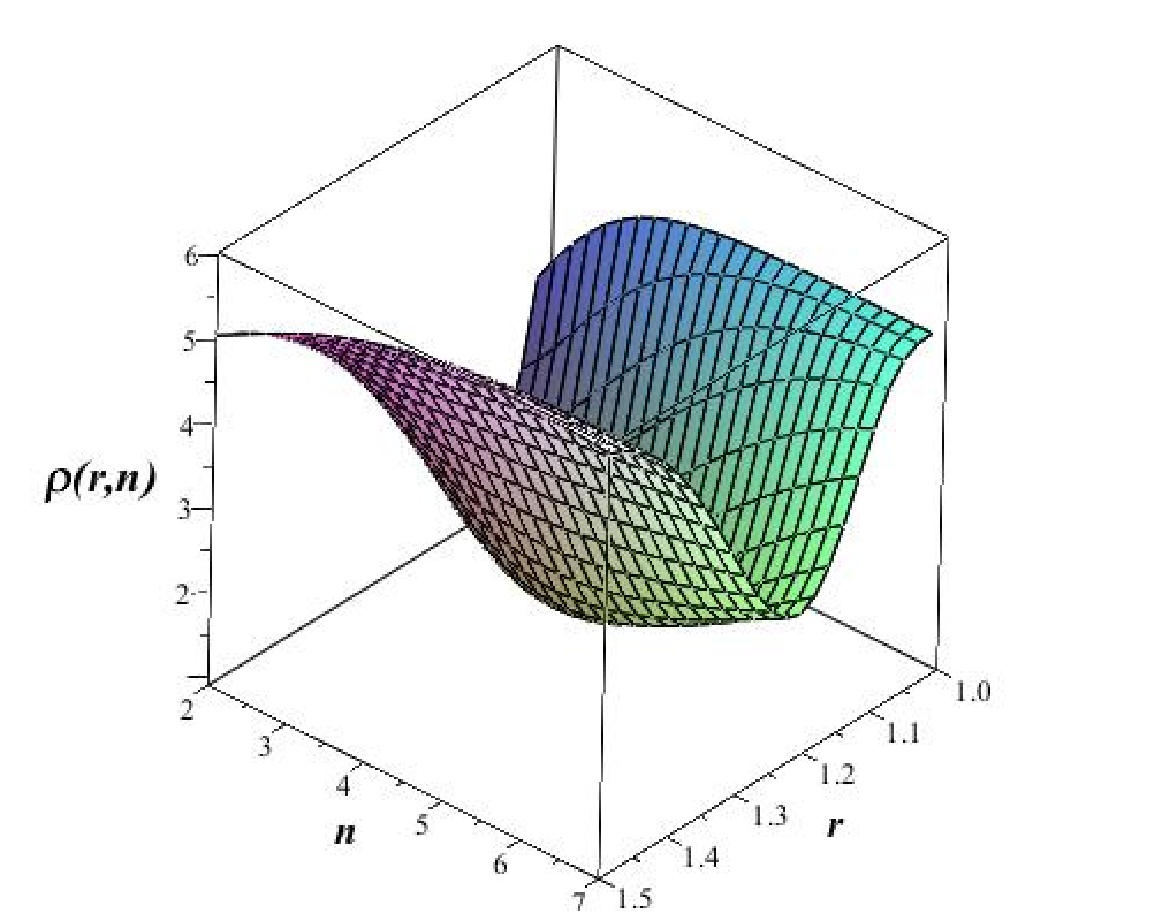}
\caption{The plot depicts the  behavior of $\rho(r,n)$ against $r$ and $n$ for the case $c=10$ and $m=-2$. One can see that $\rho(r,n)$ is a positive function in the entire range. See the text for details.}
 \label{fig6}
\end{figure}

\begin{figure}
\centering
  \includegraphics[width=3 in]{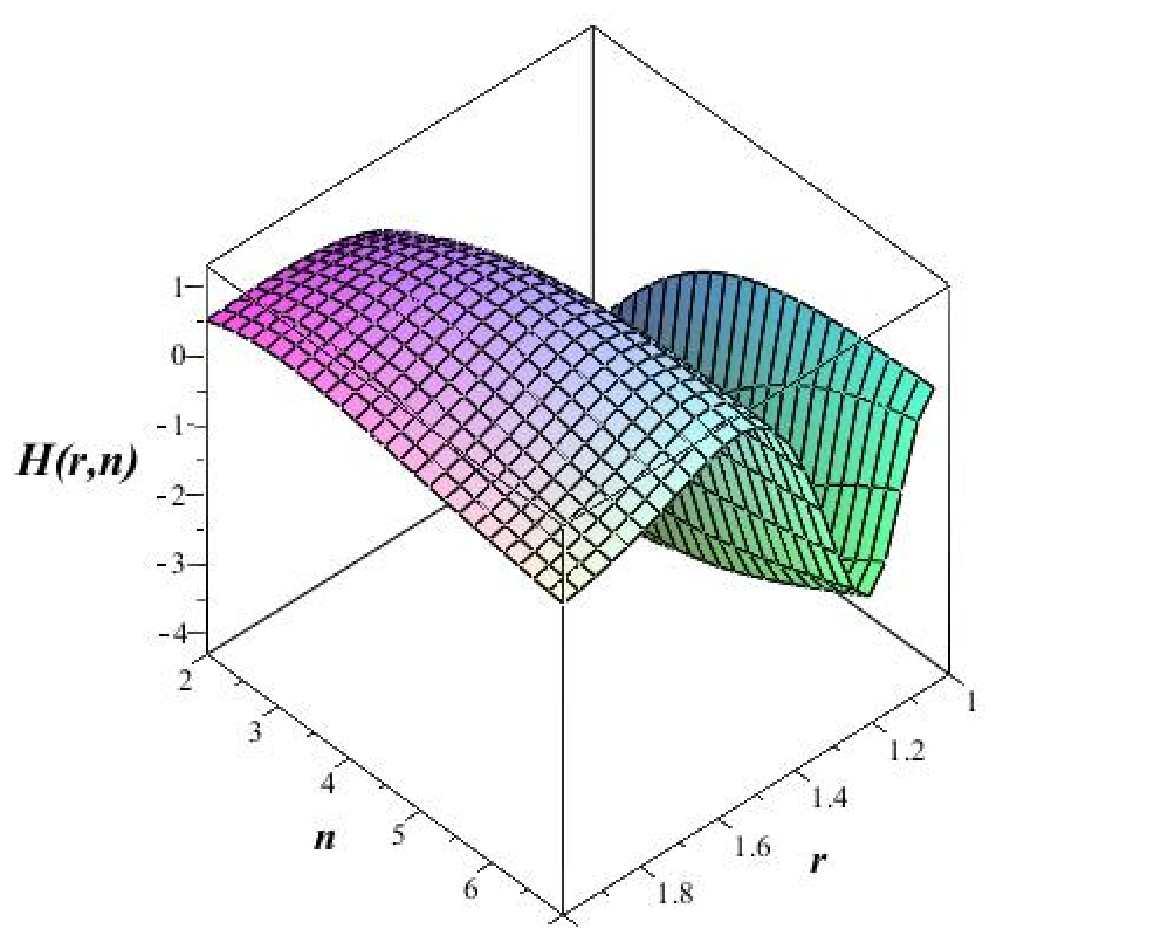}
\caption{The graphical behavior of $H(r,n)$ against $r$ and $n$ for the case $c=10$ and $m=-2$.
It shows that ECs are violated. See the text for details.}
 \label{fig7}
\end{figure}

The function (\ref{25}) is a special case of the function
\begin{equation}\label{27}
f(Q)=(-Q)^{n}+c.
\end{equation}
Studying solutions for this function in the general form is complicated in contrast to the previous functions. Let us study some samples for the free parameters $c, n$, and $m$. As the previous case, it can be concluded that $c=2 \rho_{\infty}$ so the positive $c$ is acceptable. For the next class of solutions, we will investigate solutions with a constant $c$ and variable $n$. It should be mentioned that the vanishing $c$ leads to solutions which have been studied in Sec. (\ref{subsec1}). Figure (\ref{fig6})  demonstrates that a positive energy density may be reachable for some range of $n$. To complete our study, we have plotted $H(r,n)$ for $m=-2$ in Fig.(\ref{fig7}) which shows that this class of solutions can not respect the ECs. The same result is concluded for the values $m=1/2, 0, -1/2, -1, -3, -4,-5, -8$ in the shape function. It seems that the shape function (\ref{24}) can not provide a suitable physical solution with a $f(Q)$ in the form of (\ref{27}).

\begin{figure}
\centering
  \includegraphics[width=3 in]{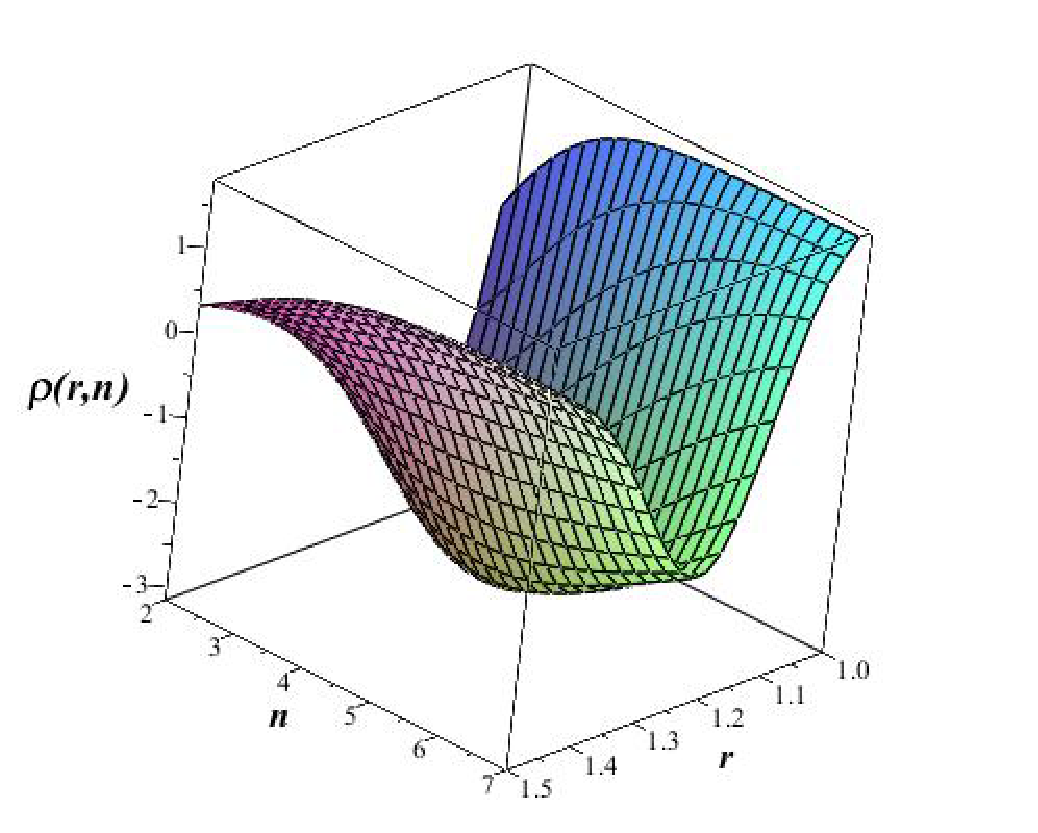}
\caption{The graphical behavior of $\rho(r,n)$  against $r$ and $n$ for the case $f(Q)=(-Q)^{n}+Q$, and $m=-2$. It shows that the energy density is negative in some range. See the text for details.}
 \label{fig8}
\end{figure}

\begin{figure}
\centering
  \includegraphics[width=3 in]{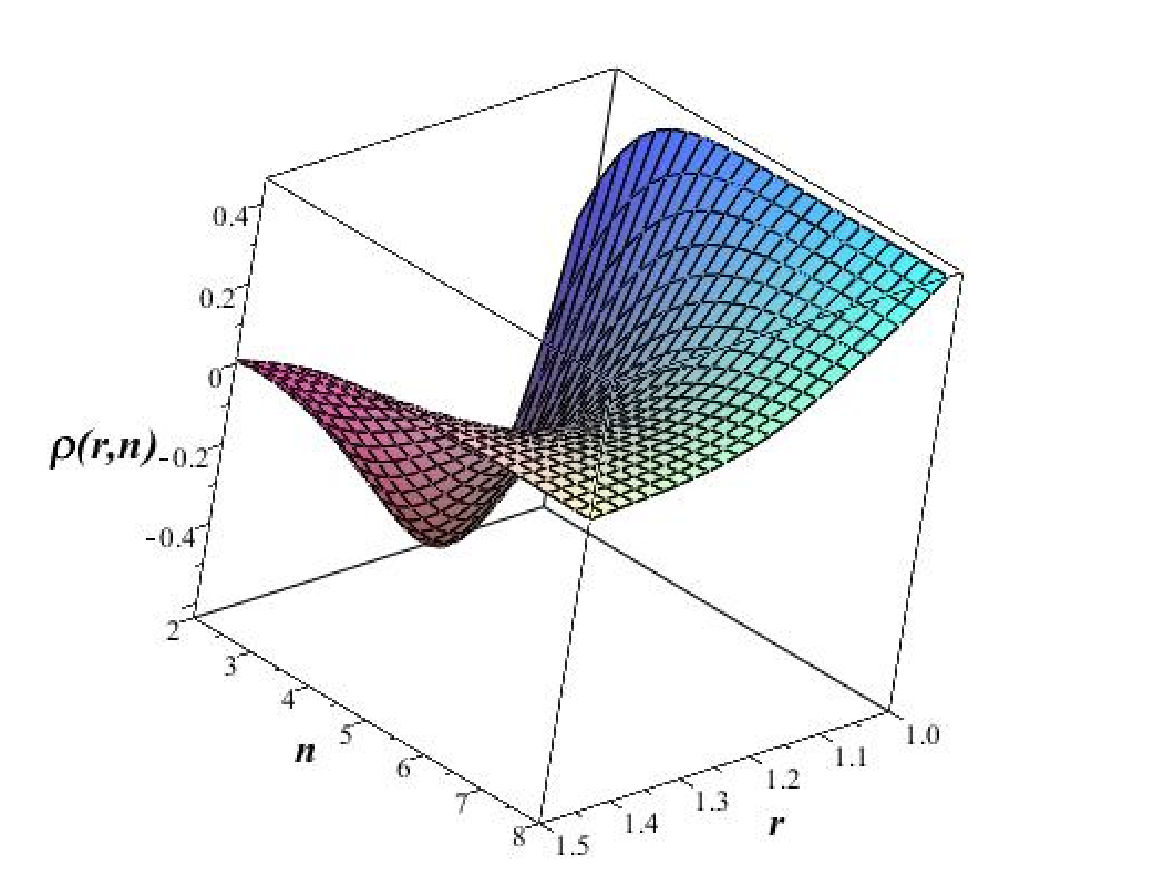}
\caption{The figure represents the $\rho(r,n)$ against radial coordinate for $f(Q)=(-Q)^{n}+Q$, and $m=-1/2$ which shows $\rho(r,n)$ is  positive in some regions. See the text for details.}
 \label{fig9}
\end{figure}

 \begin{figure}
\centering
  \includegraphics[width=3 in]{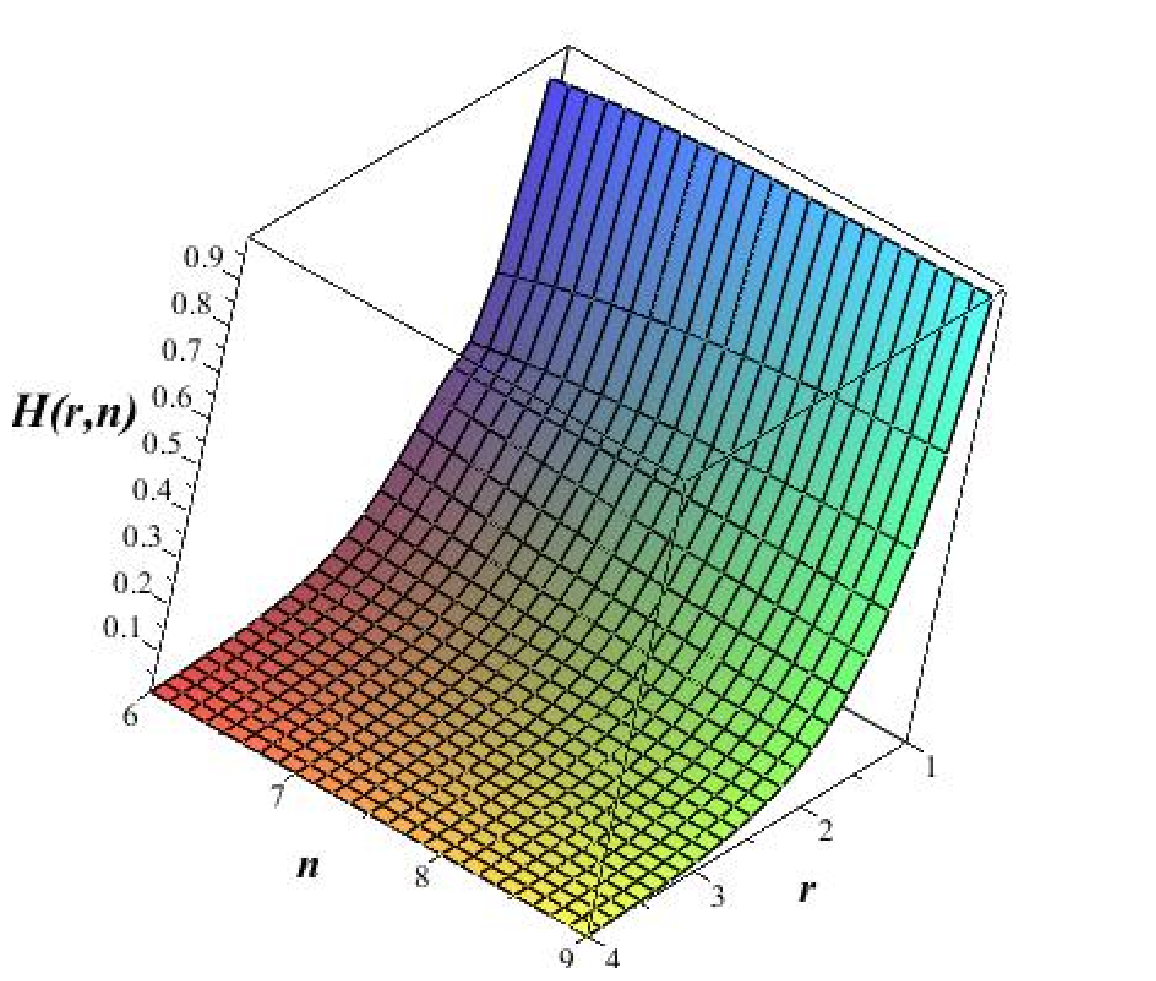}
\caption{The plot depicts  $H(r,n)$  against radial coordinate for $f(Q)=(-Q)^{n}+Q$, and $m=-1/2$ which shows $H(r,n)>0$ is reachable.  See the text for details.}
 \label{fig10}
\end{figure}

\subsection{Solutions satisfying energy conditions }\label{subsec3}
Now, let us take the $f(Q)$ function in the form
\begin{equation}\label{28}
f(Q)= (-Q)^{n}+BQ+c.
\end{equation}
It can be seen that a linear term $BQ$ is added to the previous $f(Q)$ function. In this case, the free parameters are more in contrast to the previous cases so we should fix some initial conditions to study the  solutions. It can be deduced that $c=2\rho_{\infty}$ therefore, we consider a vanishing energy density at large scale so the term $c$ is omitted. As  the first model, we consider
\begin{equation}\label{29}
f(Q)= (-Q)^{n}+Q.
\end{equation}
where $B=1$ and $c=0$ are supposed.
\begin{figure} [!h]
\centering
  \includegraphics[width=3 in]{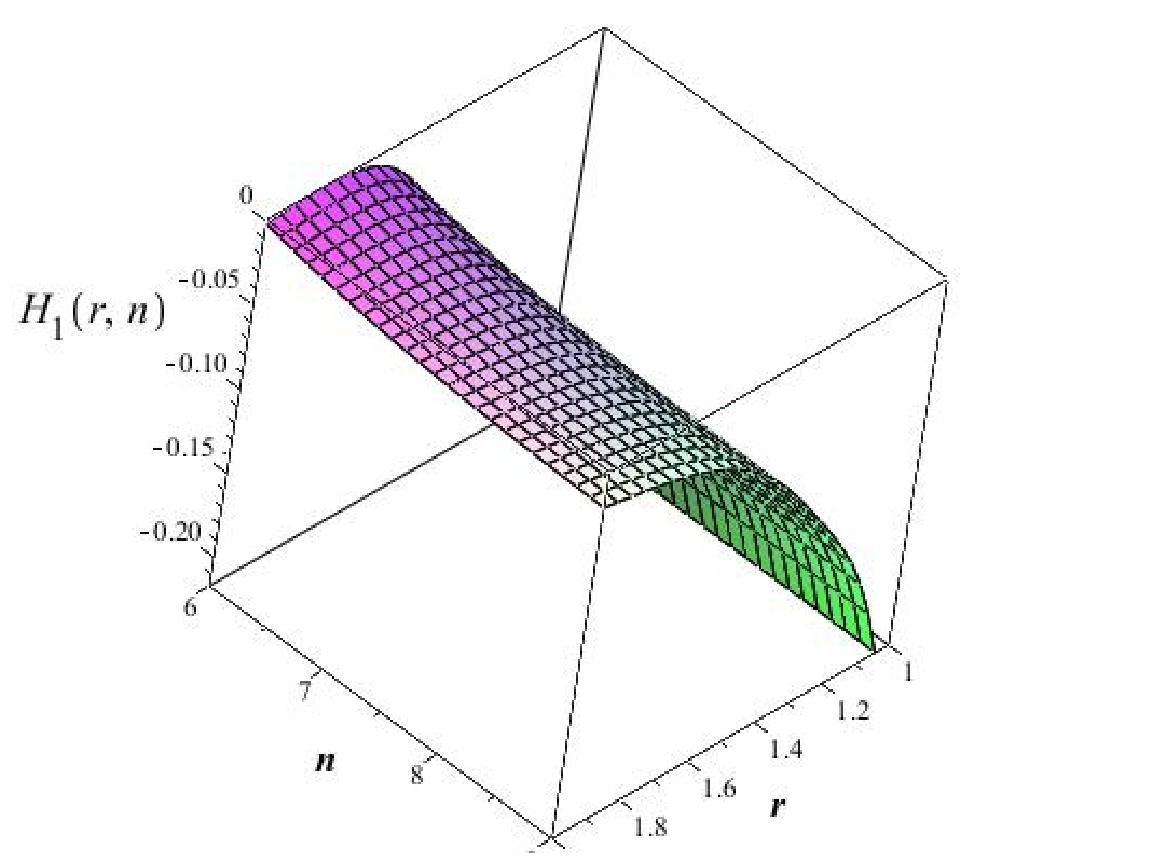}
\caption{The graphical behavior of $H_1(r,n)$ against $r$ and $n$ for the case $f(Q)=(-Q)^{n}+Q$, and $m=-1/2$. It shows that $H_1(r,n)>0$ is not reachable. See the text for details.}
 \label{fig11}
\end{figure}

\begin{figure}
\centering
  \includegraphics[width=3 in]{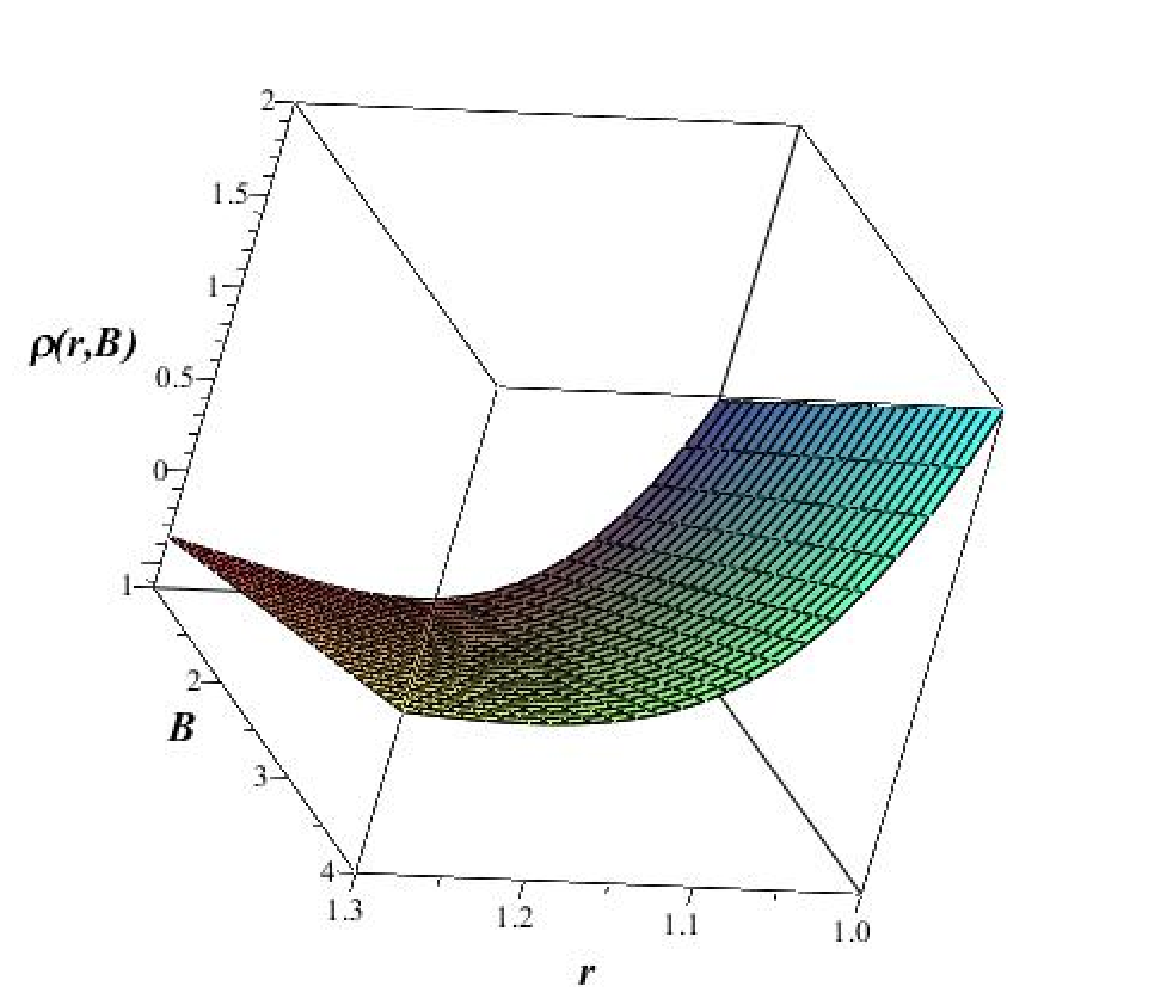}
\caption{The plot depicts  $\rho(r,B)$  against radial coordinate and $B$ for $f(Q)=Q^{2}+BQ$, and $m=-1/2$ which shows $\rho(r,B)$ is  positive in some regions of $B$. See the text for details.}
 \label{fig12}
\end{figure}
\begin{figure}
\centering
  \includegraphics[width=3 in]{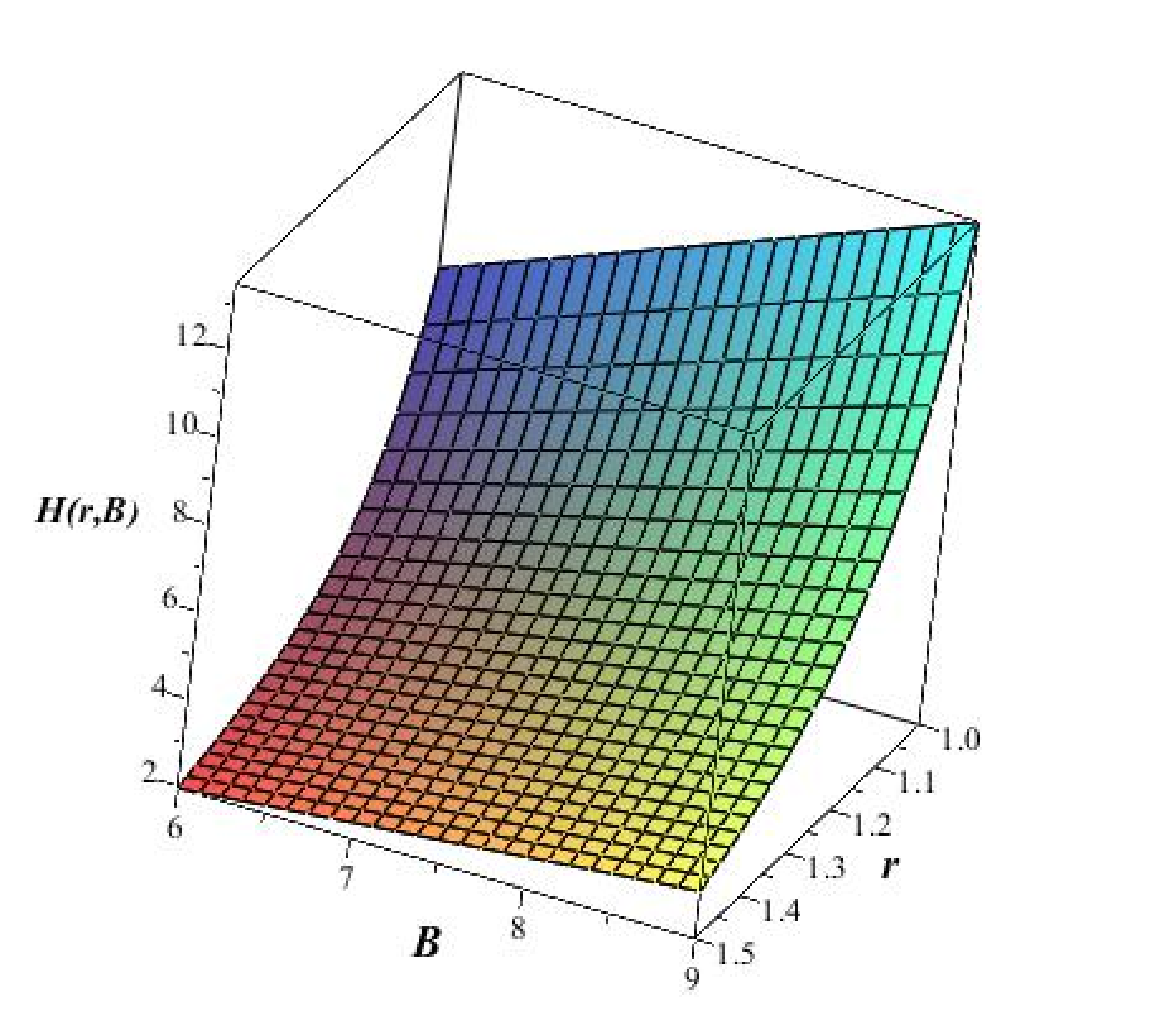}
\caption{The plot depicts  $H(r,B)$  against radial coordinate and $B$ for $f(Q)=Q^{2}+BQ$, and $m=-1/2$ which shows $H(r,B)$ is always positive.  See the text for details.}
 \label{fig13}
\end{figure}

\begin{figure}
\centering
  \includegraphics[width=3 in]{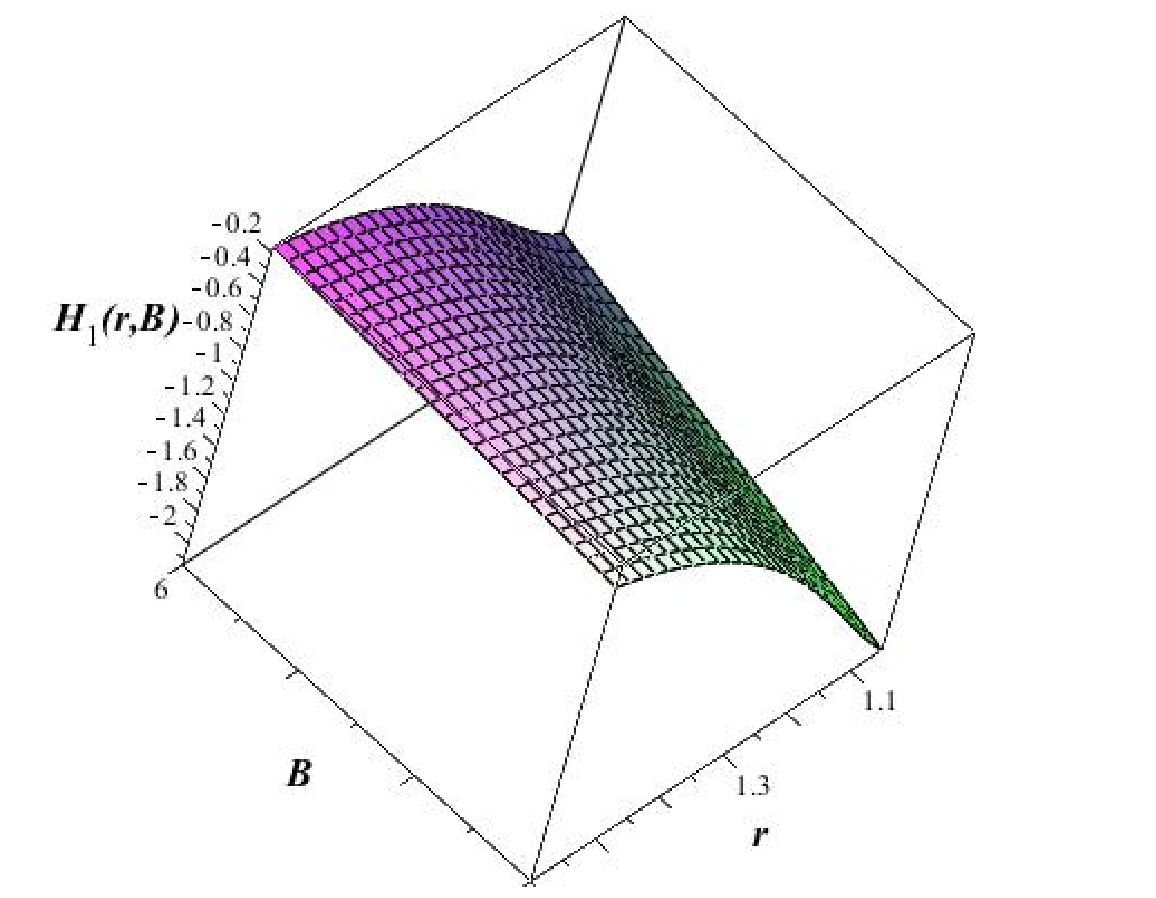}
\caption{The plot depicts  $H_1(r,B)$  against radial coordinate and $B$ for $f(Q)=Q^{2}+BQ$, and $m=-1/2$ which shows $H_1(r,B)$ is always negative.  See the text for details.}
 \label{fig14}
\end{figure}

Let us investigate this model for some special values of $m$ in the shape function. We have plotted the energy density as a function of $r$ and $n$ for $m=-2$ in Fig.(\ref{fig8}). As one can notice, the energy density is negative in some range of radial coordinate for the entire range of $n$. The same result is achieved for the values $m=-3,-4,-5,-8$ which is not physically considerable. Let us study the special case $m=-1/2$ which leads to a positive energy density for some range of $n$. In this case, we have plotted $\rho(r,n)$ as a function $r$ and $n$ in Fig.(\ref{fig9}) which shows that energy density is positive in some range of $n$. Also, the function $H(r,n)$ is plotted as a function of $r$ and $n$ in Fig.(\ref{fig10}). One can deduce from Fig.(\ref{fig10}) that the condition $H>0$ is reachable. As the next condition, the function $H_{1}(r,n)$ is depicted in Fig.(\ref{fig11}) which demonstrates that this class of solutions can not satisfy the ECs.

\begin{figure}
\centering
  \includegraphics[width=3 in]{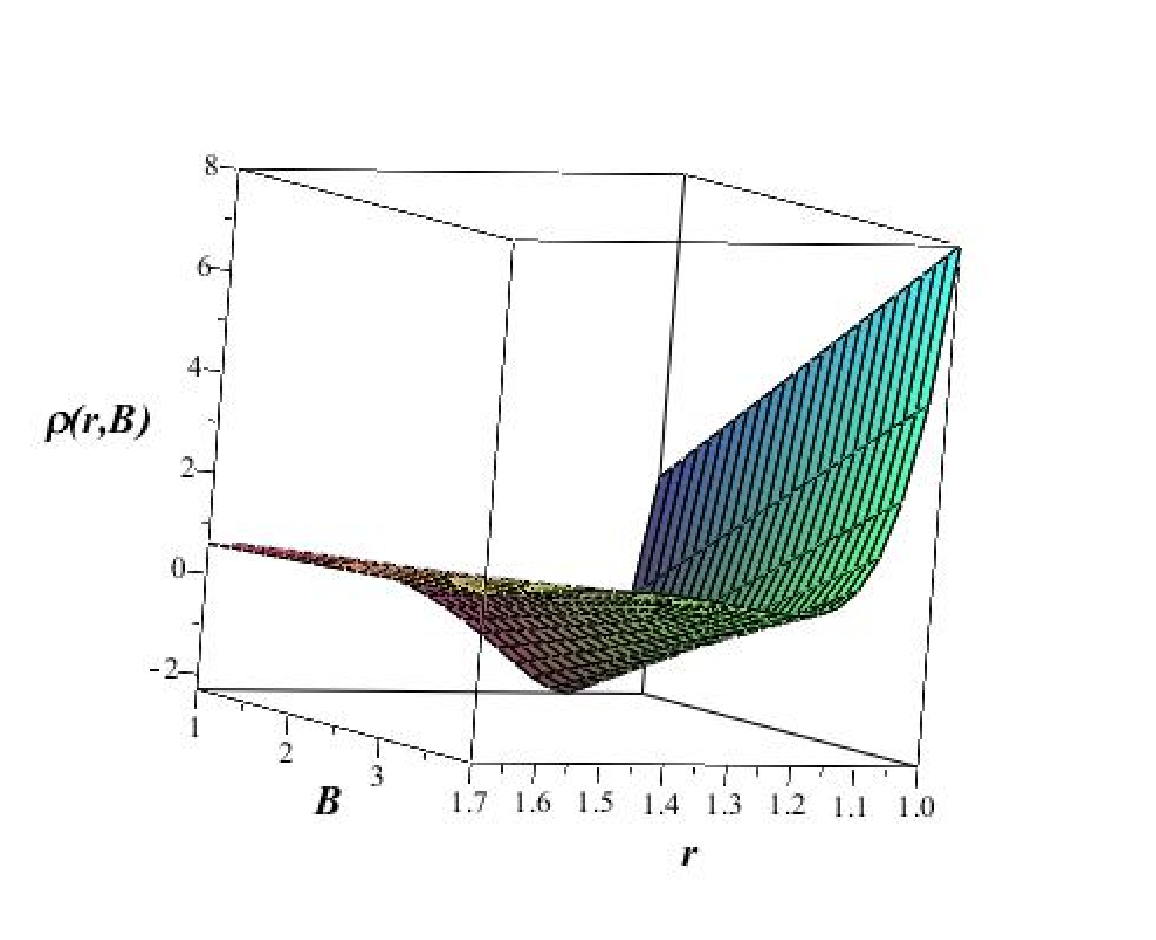}
\caption{The plot depicts  $\rho(r,B)$  against radial coordinate and $B$ for $f(Q)=Q^{2}+BQ$, and $m=-2$ which shows $\rho(r,B)$ is  positive in some regions of $B$. See the text for details.}
 \label{fig15}
\end{figure}
\begin{figure}
\centering
  \includegraphics[width=3 in]{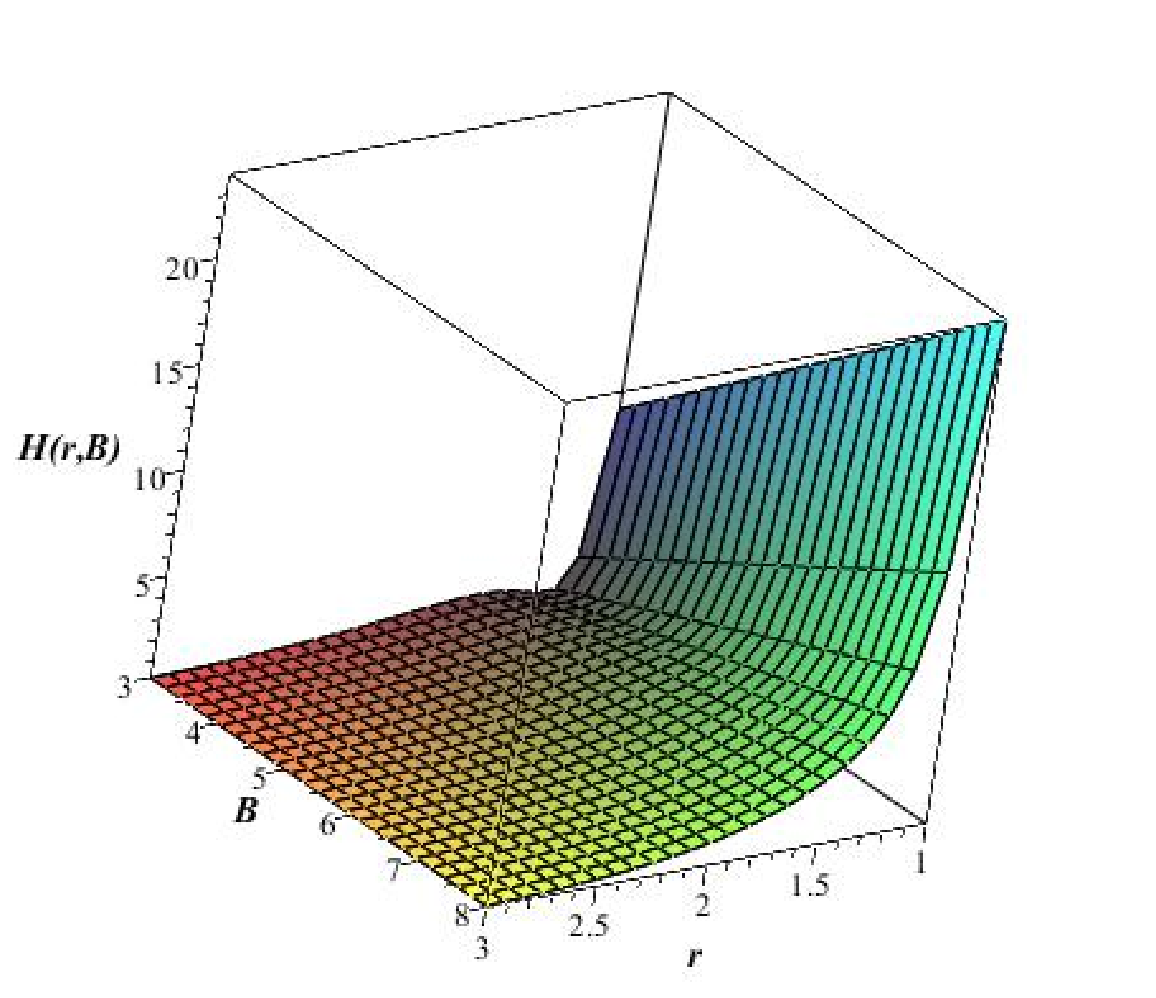}
\caption{The plot depicts  $H(r,B)$  against radial coordinate and $B$ for $f(Q)=Q^{2}+BQ$, and $m=-2$ which shows $H(r,B)$ is always positive.  See the text for details.}
 \label{fig16}
\end{figure}
  \begin{figure}
\centering
  \includegraphics[width=3 in]{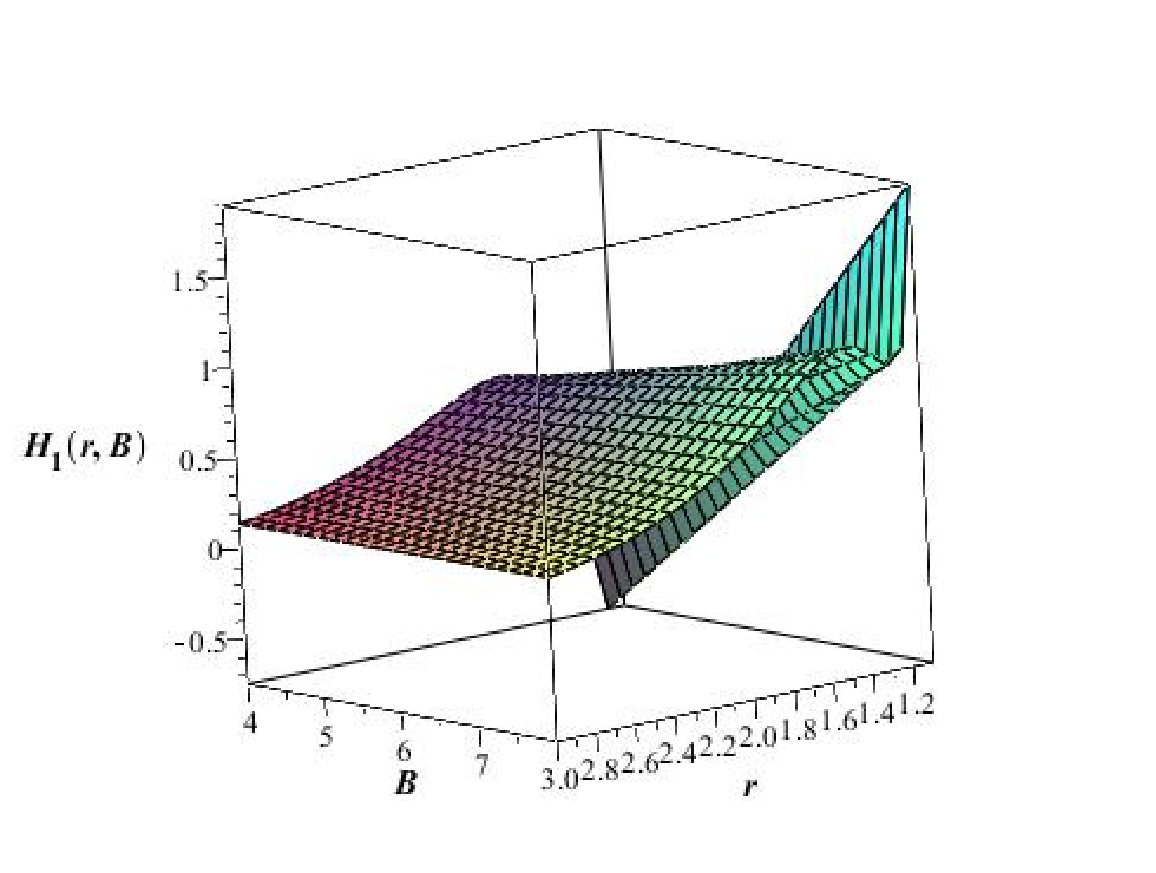}
\caption{The plot depicts  $H_1(r,B)$  against radial coordinate and $B$ for $f(Q)=Q^{2}+BQ$, and $m=-2$ which shows $H_1(r,B)$ is positive for some range of $B$.  See the text for details.}
 \label{fig17}
\end{figure}
\begin{figure}
\centering
  \includegraphics[width=3 in]{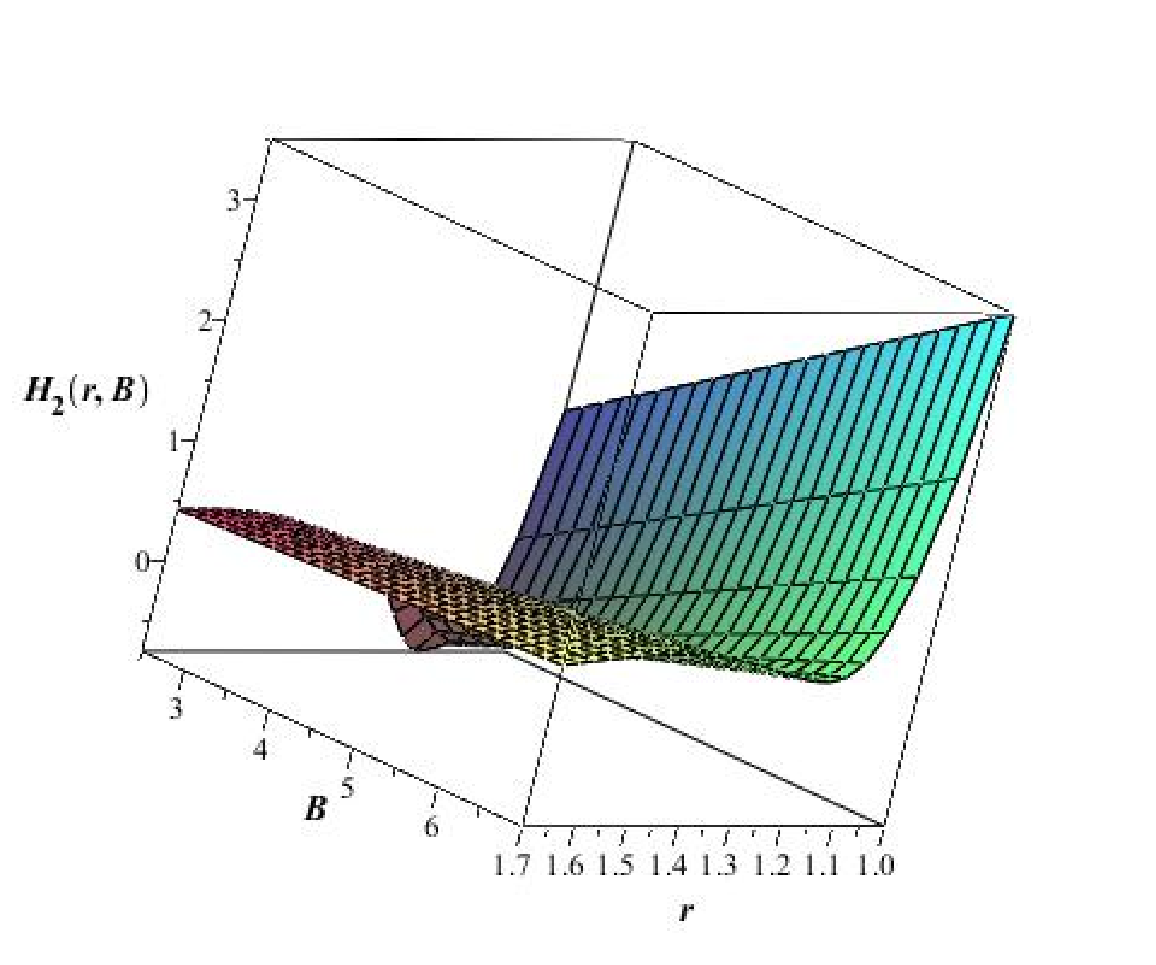}
\caption{The plot depicts  $H_2(r,B)$  against radial coordinate and $B$ for $f(Q)=Q^{2}+BQ$, and $m=-2$ which shows $H_2(r,B)$ is  positive in some regions of $B$. See the text for details.}
 \label{fig18}
\end{figure}

Due to the high complexity of the  $f(Q)$ model in (\ref{29}), we will consider
\begin{equation}\label{30}
f(Q)= (-Q)^{2}+BQ
\end{equation}
which is a special case of (\ref{29}) with $n=2$. Also, it  can be considered as a special case of (\ref{27}) while a linear term $BQ$ is added. It is easy to show that
\begin{equation}\label{31}
B=2 \rho (r_0)
\end{equation}
 In \cite{Ban}, it was shown that wormhole solutions could not exist for the specific form function (\ref{30}). But it should be mentioned that the authors in \cite{Ban} have used the  field equations, presented in \cite{Zha} instead of field equations in \cite{Mus}. In this paper, our field equations are the same as \cite{Mus}. In the next step, we will investigate the essential conditions to respect the ECs. As the first case, we have chosen $m=-1/2$ in the shape function then we have plotted the energy density as a function of $r$ and $B$ in Fig.(\ref{fig12}). This figure shows that $\rho>0$ is reachable for some range of $B$. We also found that $H>0$ is reachable for some range of $B$ from Fig.(\ref{fig13}). For checking the condition $H_{1}>0$, we have plotted $H_{1}$ as a function of $r$ and $B$ in Fig.(\ref{fig14}) which presents an unsatisfactory result.

\begin{figure}
\centering
  \includegraphics[width=3 in]{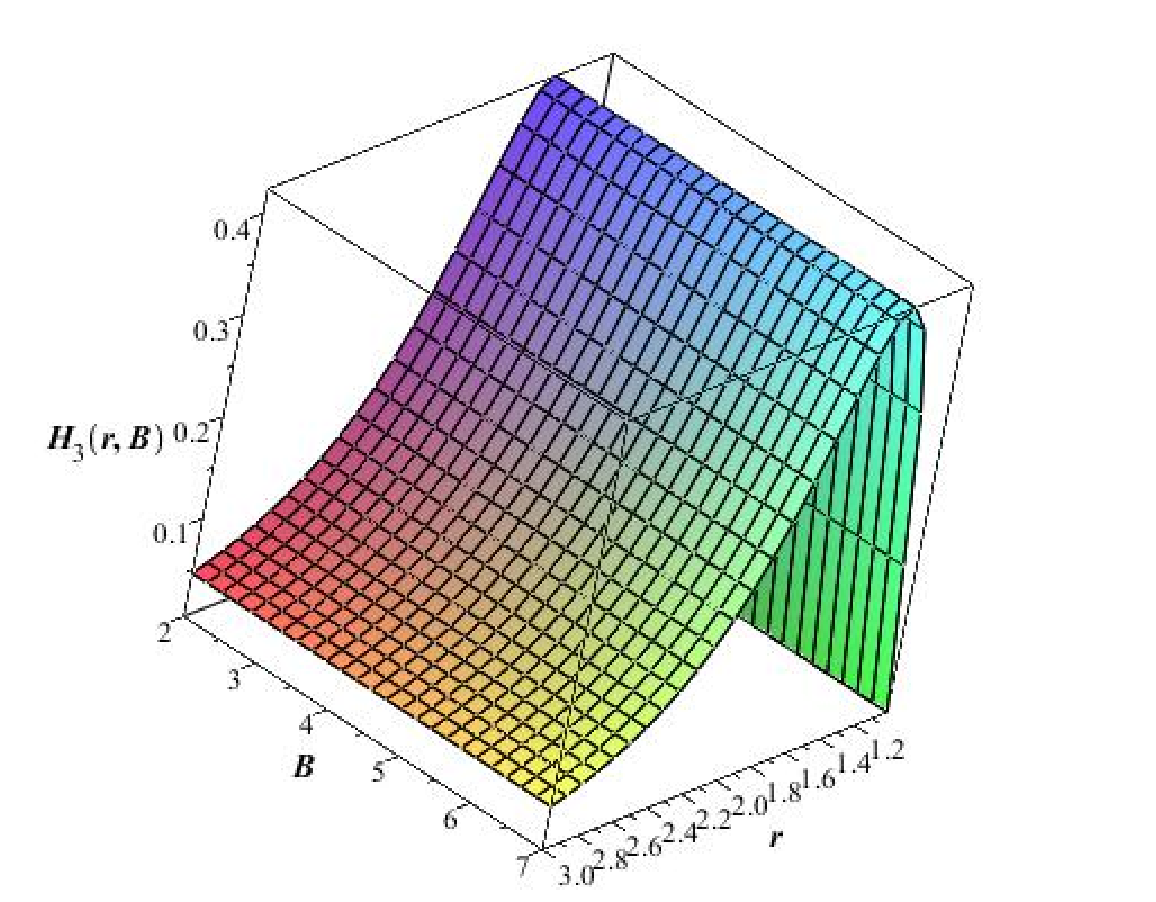}
\caption{The plot depicts  $H_3(r,B)$  against radial coordinate and $B$ for $f(Q)=Q^{2}+BQ$, and $m=-2$ which shows $H_3(r,B)$ is always positive.  See the text for details.}
 \label{fig19}
\end{figure}

\begin{figure}
\centering
  \includegraphics[width=3 in]{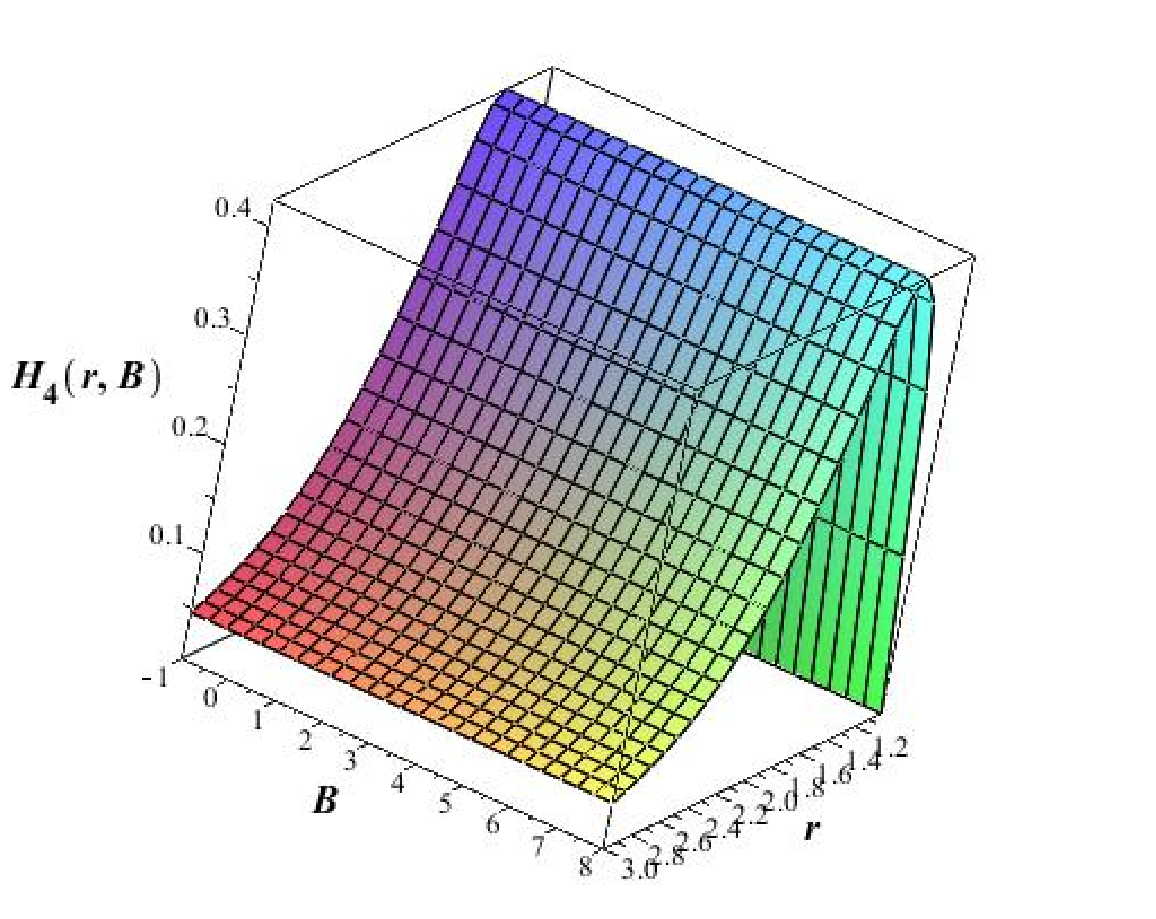}
\caption{The plot depicts  $H_4(r,B)$  against radial coordinate and $B$ for $f(Q)=Q^{2}+BQ$, and $m=-2$ which shows $H_4(r,B)$ is always positive . See the text for details.}
 \label{fig20}
\end{figure}

 Now, we will study the form function (\ref{24}) for some other values of $m$. Our investigations show that the results for $m=1/2, -1/2, -1$ are unsatisfactory but for $m=-2$ are significate. We would like to clarify the results for the case $b(r)=1/r^{2}$ and $f(Q)=Q^{2}+BQ$ in the recent part of this section. Figure (\ref{fig15}) demonstrates the behavior of energy density as a function of $r$ and $B$. One can conclude from Fig.(\ref{fig15}) that $\rho>0$ is reachable for some value of $B$. In Fig.(\ref{fig16}), we have depicted the $H(r,B)$ as a function of $r$ and $B$ which implies $H>0$ is reachable.
  The conditions $H_{1}>0$, $H_{2}>0$, $H_{3}>0$ and $H_{4}>0$ are also reachable (see Figs.(\ref{fig17} - \ref{fig20})). So, we have found the solution which satisfies all of the ECs for some range of $B$. It is easy to show that
  \begin{equation}\label{32}
\omega_{\infty}=\lim_{r\longrightarrow \infty}\frac{p}{\rho}=\frac{1}{7},
\end{equation}
which shows the EoS is asymptotically linear \cite{variable}
   \begin{equation}\label{32a}
p(r)=(\omega_{\infty}+g(r))\rho (r).
\end{equation}
It should be noted that
  \begin{equation}\label{32b}
\lim_{r\longrightarrow \infty}g(r)=0.
\end{equation}
One can see that the this class of solutions are not isotropic.

   Let us investigate the effect of the free parameters in (\ref{28}) on the  wormhole solutions. The results for some individual $B$ and $m$ parameters are presented in tables (\ref{fe5}) and (\ref{t55}) . These results have allowed us to give a better understanding of the influence of the free parameters in $f(Q)$ function and shape function on the validation of ECs. It is clear that $B$ has a crucial role in the satisfaction of ECs. Table (\ref{fe5}) indicates that in the case $B=3$, none of the presented individual parameters for $n$ can  provide a solution  that satisfies the ECs. But in the case $B=5.5$, some ECs are satisfied for $n=3/2,2$. It seems that  the most of individuals $n$ respect the ECs while $B$ increases. One also may see that the case $n=2/3$ violates all of the ECs independent of the value of $B$. Generally, due the non-linearity of the field equation in the $f(Q)$ scenario, we can not introduce any explicit relation between the  validation of the ECs and the role of $n$ and $B$.

 All of the $f(Q)$ models which have been studied in this paper, are some special cases of a general form
  \begin{equation}\label{33}
f(Q)=D(-Q)^{n}+BQ+c.
\end{equation}
It is easy to check that for $n=0$ and $B=1$, the model reduces to the standard  symmetric teleparallel equivalent of general relativity  model with the quantity $(D+c)/2$ playing the role of the cosmological constant \cite{Tel, Tel2, Ec}. On the other hand,  the symmetric teleparallel equivalent of GR can be achieved  while $n=1$ is considered. It was shown that modification
from the GR evolution occurs at low curvatures regime for $n < 1$ and occurs at high curvatures regime  for $n > 1$. Hence, it can be deduced that $n > 1$ will be applicable for the early Universe, while $n < 1$ will be applicable to the late-time DE dominated Universe \cite{Khy}.

 However, for the sake of completeness, we continue to investigate some other features of the free parameters in this model for wormhole theory. Thus, in summary, we have presented the results of some individual values for $B$ and $D$ while $b(r)=\frac{1}{r^{2}}$ and $n=2$ are considered (Table(\ref{t55})). Remarkably, we have found that the existence of an extra parameter like $D$ can change at least the limits of $B$ for validations of the ECs. For instance, for the case $B=3$, one can find that there are not any solutions that respect the ECs in the case $D=1$ but the case $D=1/2$ provides this possibility. Although the exact relations between the influence of the free parameters in (\ref{33})  on validation of ECs can not be addressed but one can deduce from the Tables (\ref{fe5}) and (\ref{t55}) that the changes in the free parameters affected directly  the ECs.

In \cite{Ec}  a complete test of ECs for some $f(Q)$  gravity models is presented. It was shown that the ECs allowed us to fix our free parameters, restricting the families of $f(Q)$ models compatible with the accelerated expansion of the Universe passes through. Mandal et al. have shown that the ECs directly depend on the free parameters $D$ and $n$ so one cannot take these values  arbitrarily, which may violate the ECs as well as the current scenario of the Universe dominated by the dark energy \cite{Ec}.

\begin{table*}[]
\begin{tabular}{|l|l|l|l|l|l|l|l|l|}
\hline
$B$ & $n$ & $\rho>0$ & $H>0$  & $H_1>0$ & $H_2>0$ & $H_3>0$ & $H_4>0$  & Satisfied ECs  \\ \hline
 3 & $ \frac{2}{3}$ & $\times$ & $\times$ & $\times$ & $\times$ & $\times$ & $\times$ &   \\ \hline
 3 & $ \frac{3}{2}$  & \checkmark & \checkmark & $\times$ & $\times$ & $\times$   & \checkmark  & \\ \hline
 3 &  2 & $\times$ & \checkmark &  $\times$& $\times$ & $\times$ & \checkmark &  \\ \hline
 3 &  3 & $\times$ &$\times$  & $\times$ & $\times$ &$\times$ & \checkmark &  \\ \hline
 3 &  6 & $\times$ & \checkmark & $\times$ & $\times$ & $\times$ & \checkmark  &  \\ \hline
 5.5 &  $ \frac{2}{3}$ & $\times$ & $\times$  & $\times$  & $\times$  & $\times$ & $\times$ &  \\ \hline
 5.5 & $ \frac{3}{2}$  & \checkmark & \checkmark & \checkmark & \checkmark & \checkmark  & \checkmark &  $NEC$,$WEC$,$DEC$, $SEC$ \\ \hline
 5.5 & 2  & \checkmark & \checkmark & \checkmark & $\times$ & \checkmark  & \checkmark &  $NEC$,$WEC$, $SEC$ \\ \hline
 5.5 & 3  & \checkmark & \checkmark & $\times$ & $\times$ &  $\times$ & \checkmark & \\ \hline
 5.5 &  6 & \checkmark & \checkmark & $\times$ & $\times$ &  $\times$ & \checkmark & \\ \hline
 10 & $ \frac{2}{3}$  & $\times$ & $\times$  & $\times$  & $\times$  &  $\times$  & $\times$ & \\ \hline
 10 & $ \frac{3}{2}$  & \checkmark & \checkmark & \checkmark & \checkmark & \checkmark  & \checkmark &$NEC$,$WEC$,$DEC$, $SEC$ \\ \hline
 10 & 2  & \checkmark & \checkmark & \checkmark & \checkmark & \checkmark  & \checkmark & $NEC$,$WEC$,$DEC$, $SEC$ \\ \hline
 10 & 3  & \checkmark & \checkmark & \checkmark & \checkmark &  \checkmark & \checkmark& $NEC$,$WEC$,$DEC$, $SEC$ \\ \hline
 10 &  6 & \checkmark & \checkmark & \checkmark & \checkmark & \checkmark  &\checkmark & $NEC$,$WEC$,$DEC$, $SEC$ \\ \hline
\end{tabular}
\caption{The results for some individual $B$ and $n$ in  $f(Q)=(-Q)^{n}+BQ$ while $b(r)=\frac{1}{r^{2}}$  is considered.  }\label{fe5}
\end{table*}

\begin{table*}[]
\begin{tabular}{|l|l|l|l|l|l|l|l|l|}
\hline
$B$ & $D$ & $\rho>0$ & $H>0$  & $H_1>0$ & $H_2>0$ & $H_3>0$ & $H_4>0$  & Satisfied ECs  \\ \hline
 3 & $ -1$ & $\times$ & $\times$ & $\times$ & $\times$ & $\times$ & $\times$ &   \\ \hline
 3 & $ \frac{1}{2}$  & \checkmark & \checkmark & \checkmark & \checkmark & \checkmark   & \checkmark  & $NEC$,$WEC$,$DEC$, $SEC$ \\ \hline
 3 &  1 & $\times$ & \checkmark &  $\times$& $\times$ & $\times$ & \checkmark &  \\ \hline
 3 &  2 & $\times$ &$\times$  & $\times$ & $\times$ &$\times$ & \checkmark &  \\ \hline
 3 &  6 & $\times$ & $\times$ & $\times$ & $\times$ & $\times$ & \checkmark  &  \\ \hline
 5.5 &  $-1$ & $\times$ & $\times$  & $\times$  & $\times$  & $\times$ & $\times$ &  \\ \hline
 5.5 & $ \frac{1}{2}$  & \checkmark & \checkmark & \checkmark & \checkmark & \checkmark  & \checkmark &  $NEC$,$WEC$,$DEC$, $SEC$ \\ \hline
 5.5 & 1  & \checkmark & \checkmark & \checkmark & $\times$ & \checkmark  & \checkmark &  $NEC$,$WEC$, $SEC$ \\ \hline
 5.5 & 2  & $\times$ & \checkmark & $\times$ & $\times$ &  $\times$ & \checkmark & \\ \hline
 5.5 &  6 & $\times$ & $\times$ & $\times$ & $\times$ &  $\times$ & \checkmark & \\ \hline
 10 & -1  & $\times$ & $\times$  & $\times$  & $\times$  &  $\times$  & $\times$ & \\ \hline
 10 & $ \frac{1}{2}$  & \checkmark & \checkmark & \checkmark & \checkmark & \checkmark  & \checkmark &$NEC$,$WEC$,$DEC$, $SEC$ \\ \hline
 10 & 1  & \checkmark & \checkmark & \checkmark & \checkmark & \checkmark  & \checkmark & $NEC$,$WEC$,$DEC$, $SEC$ \\ \hline
 10 & 2  & \checkmark & \checkmark & $\times$ & $\times$ &  $\times$ & \checkmark &   \\ \hline
 10 &  6 & $\times$ & $\times$ & $\times$ & $\times$ & $\times$  &\checkmark &  \\ \hline
\end{tabular}
\caption{ The results for some individual $B$ and $D$ in  $f(Q)=D(-Q)^{2}+BQ$ while $b(r)=\frac{1}{r^{2}}$  is considered.  }\label{t55}
\end{table*}

\section{Concluding remarks}

GR is the basic theory for the Standard Model of physical cosmology. Despite the  success of GR in describing many cosmological phenomena, this theory
has some limitations in characterizing some other phenomena of the cosmos. Accelerated expansion of the Universe, cosmological constant problem, coincidence problem, and  very early universe  are some examples \cite{Pro}. The modifications of gravity are proposed as an alternative to investigating these problems. Modified theories of gravity also illustrate the problems faced by models which favor goodness of fit over parsimony. The traversable wormholes violate ECs in the context of GR. Although there are not any clear or verified  experimental results which establish the existence of the wormhole, violation of ECs in standard GR is an important task that permits us to test the other modified theory to solve this problem. Investigation of the wormhole in the modified gravities may open new windows to observe or construct a wormhole. Discovering exact wormhole solutions represents a pivotal aspect of wormhole research, with the most crucial challenge lying in the exotic matter involved.

 Various techniques have been employed in existing literature to identify exact wormhole solutions, some proposing approaches to minimize the reliance on exotic matter. Additionally, researchers have uncovered wormholes which respect the ECs in the framework of modified gravity theories.
 Recently a new proposal teleparallel symmetric equivalent of general relativity has been used to describe wormhole solutions. This new formalism is considered as the third equivalent formulation of GR by means of the Q-scalar motivates novel ways of modifying gravity. In this work, we have studied the  implications of the new type of modified gravity theories $f(Q)$ to find solutions that satisfy ECs. Since the most of solutions which have been presented before do not respect ECs \cite{Wang, Sha, Mus, Hassan, sym, Ban, Calz, f(Q), Kir}, our solutions seem to be new and significate.

 Finding asymptotically exact wormhole solutions in the context of GR with a known EoS is not a simple task. Due to the higher order curvature terms, this procedure is more complicated in the background of $f(Q)$ gravity. Because of the aforementioned reason, we have focused on finding wormhole solutions by using some basic models of the $f(Q)$ function and a well-known shape function. The power-law shape function has been used extremely in studying wormhole solutions and is the most famous solution in the different gravity scenarios. These reasons are the motivation behind the choice of the shape function. In the second step, by carefully testing some specific $f(Q)$ functions with free parameters, the desired results have been found for some special cases of free parameters.

   The nonlinearity order of field equations in various modified theories increases the complexity of wormhole solutions. Using the fine-tuning technique for free parameters can reduce this complexity. This technique admits us to discover solutions which satisfy ECs  for some free parameters. The results are strongly depend on the numerical values of the model parameters. It was shown that $B$ which has a crucial role in ECs  can be related to the values of the energy momentum tensor at the throat of the wormhole. Also, we have shown that  the constant parameter $c$ in  $f(Q)$ model must be interpreted as the energy density at the large scale. We have omitted this parameter to have more viable solutions.
  On the other hand, the boundary conditions have been investigated to improve the viability of solutions.  Furthermore, we have also explored the EoS for our solutions which verified that an anisotropic matter content  is essential to sustain wormholes in this realm. It was shown that the EoS is asymptotically linear.

To summarize, the violation of ECs is a fundamental inconsistency that needs to be addressed in wormhole theory. By fine-tuning the free parameters in the shape function and $f(Q)$ models, we have found solutions which require no-exotic matter. So, this is the significant point in this work where exotic matter is just replaced by an equivalent modified form of gravity. Although, we have shown the presence of $f(Q)$ gravity will be enough to sustain a traversable wormhole without exotic matter the viability of these models should be tested more precisely in a cosmological background. Using these results  with astronomical results in the context of $f(Q)$ gravity can provide the best proposal for $f(Q)$ models or at least the ability of each model to describe cosmological phenomena. These solutions may represent an alternative to the standard GR scenario.

 Although wormholes have not been detected experimentally  yet, in this study, we have  explored  the possible existence of some  wormhole geometries in the context of $f(Q)$ gravity. The theoretical consistency and motivations on these extensions of $f(Q)$ can be established to explore the new avenues in the wormhole theory and cosmological predications of $f(Q)$ theory.  Along this way, we have considered a vanishing  redshift function, i.e., $\phi(r) = 0$, but solutions with non-constant redshift function can be explored. Furthermore, our algorithms can be used for some other models of the $f(Q)$ or different forms of shape function.

\end{document}